\documentclass[oneside,english]{aa}
\usepackage[T1]{fontenc}
\usepackage[latin1]{inputenc}
\usepackage{babel}
\makeatletter

\providecommand{\LyX}{L\kern-.1667em\lower.25em\hbox{Y}\kern-.125emX\@}

\usepackage{txfonts}

\usepackage{natbib}
\bibpunct{(}{)}{;}{a}{}{,}

\usepackage{graphicx}
\usepackage{amstext}
\usepackage{amscd}

\renewcommand{\cite}[2][]{\citep[#1]{#2}}
\newcommand{\kommentar}[1]{}
\newcommand{\Ableitung}[2]{\frac{\mathrm{d}#1}{\mathrm{d}#2}}
\newcommand{\Partiell}[2]{\frac{\partial #1}{\partial #2}}
\renewcommand{\text}[1]{\mathrm{#1}}
\newcommand{\tr}{\mathrm{tr}\,}
\renewcommand{\Re}{\mathrm{Re}\,}
\newcommand{\corr}[1]{#1}
\makeatother
\begin{document}
\kommentar{

\newcommand{\Ableitung}[2]{\frac{d#1 }{d#2 }}

\newcommand{\Partiell}[2]{\frac{\partial #1 }{\partial #2 }}

}

\titlerunning{Long-term evolution of compact binaries \ldots}

\abstract{We resume the discussion about irradiation-driven mass transfer
cycles in semi-detached compact binary systems. The analytical model
that describes the onset of these cycles, which occur on a thermal
timescale of the donor star, is reexamined. We take into account a
contribution of the thermal relaxation which is not related to the
irradiation of the donor star and which was neglected in previous
studies. Cataclysmic variables (CVs) containing extended giant donors
are more stable than previously thought. CVs close to the upper edge
of the period gap can undergo cycles for low angular momentum loss
rates, as they have been suggested by recent magnetic braking prescriptions,
while they are stable for high braking rates.

A model for the irradiation geometry that takes into account surface
elements near the terminator of the donor star indicates that possibly
also low-mass X-ray binaries (LMXBs) can undergo mass transfer cycles.
Regarding the braking rate, which is necessary to drive cycles, basically
the same restrictions apply for short period LMXBs as they do for
short period CVs. We confirm that LMXBs containing giants can undergo
cyles. In terms of an irradiation efficiency parameter \( \alpha  \)
CVs are susceptible to the irradiation instability for \( \alpha \gtrsim 0.1 \)
while LMXBs are susceptible for \( \alpha \lesssim 0.1 \).

The predictions of the analytical model are checked by the first long-term
evolutionary computations of systems undergoing mass transfer cycles
with full 1D stellar models. For unevolved main sequence (MS) and
giant donors the analytic model provides reasonable values for the
boundaries of the stable and unstable regions while CVs containing
highly evolved MS donors are more stable at high braking rates than
expected.

Taking into account irradiation the minimum period of CVs is increased
by up to 1-2 minutes, depending on \( \alpha  \).

}

\title{Long-term evolution of compact binaries \\ with irradiation feedback}

\author{Andreas B\"uning\and Hans Ritter}

\institute{Max-Planck-Institut f\"ur Astrophysik, Karl-Schwarzschild-Str. 1,
D-85740 Garching, Germany}

\offprints{Hans Ritter, \\\email{hsr@mpa-garching.mpg.de}}

\date{received/accepted}

\maketitle

\section{Introduction}

Observations of close binary systems have shown that a star which
is illuminated by its companion can exhibit a hotter, illuminated,
and a cooler, unilluminated side  \citep[see, e.g.,][~and references therein]{Eddington26, Ritter00}.
This is the well-known reflection effect  \citep[for a review see][]{Vaz85}.
If the illuminated star has a convective envelope, then irradiation
inhibits the energy transport through the illuminated outer layers
\cite{Nordlund85}. In the case of a semi-detached compact binary
system the reflection effect can be caused by accretion luminosity
that is released nearby the compact star. In this case the energy
transport, i.e., the intrinsic flux through the illuminated outer
layers of the donor star is coupled to the mass transfer rate. This
leads to a feedback, the consequences of which are the subject of
this paper.

\corr{The influence of irradiation onto mass transfer was first discussed
by  \citet{Podsiadlowski91} for the simplified
case of symmetrical irradiation of the donor. Later, 
\citet{Hameury93} tried to simulate asymmetrical irradiation by time-periodic
symmetric irradiation and concluded that a more realistic model is
required.}

\corr{In a series of papers  \citet{Ritter95, Ritter96b, King95b, King96}
developed a feedback model for asymmetrical irradiation} that predicts
under which conditions the secular mass transfer rate becomes \corr{unstable
and} the system undergoes mass transfer cycles which occur on a thermal
timescale of the convective envelope of the donor star.

We reexamine the above-mentioned irradiation feedback model and take
into account a contribution of the thermal relaxation which is not
related to the irradiation of the donor star. It turns out that this
term, which was neglected in previous studies \corr{of the above-mentioned
analytical model}, can become important, especially for giant donors.

Unfortunately, there have been basically no numerical computations
to test the validity of this model and especially the boundaries of
the stable and unstable regions. Only few computations have been carried
out by using bipolytrope stellar models and in most cases also by
using simplified models for the treatment of irradiation \cite{Ritter95, Frank98, Ritter00}.
There has been only one short-term evolution using full 1D stellar
models and a realistic treatment of irradiation which was published
by  \citet{Hameury97}. It has served more
as a {}``proof of concept{}'' for their model of the reduction of
the intrinsic flux by illumination than as a numerical confirmation
of the irradiation feedback model.

The main purpose of this paper is to confirm the predictions of the
irradiation feedback model by numerical long-term evolutionary computations
that are based on full 1D stellar models for the donor star, a realistic
model for the irradiation geometry, and tabulated results of 
\citet{Hameury97} for the reduction of the intrinsic flux by illumination.

In Sect.~\ref{Chap:InputPhysics} we mention the input physics that
is necessary for the feedback model and for the numerical computations.
This includes the mass loss rate prescription, the reflection effect,
the irradiation geometry, and the thermal relaxation of the donor
star. Afterwards, in Sect.~\ref{Chap:AnalyticalModel} we develop
two ordinary differential equations which describe the system adequately
for our purpose, and in Sect.~\ref{Chap:LinearStability} we derive
the conditions which describe the onset of instability by means of
linear stability analysis.

We give a short overview in Sect.~\ref{Chap:Code} about input physics,
which we use in our code, and subsequently, we present our numerical
results for CVs and LMXBs containing evolved and unevolved donor stars
in Sect.~\ref{Chap:Numerics}.

\section{Input physics}

\label{Chap:InputPhysics}

We consider a semi-detached compact binary system. The compact primary
star, a white dwarf, a neutron star, or a black hole, is denoted by
subscript \( 1 \), the secondary star, a main sequence star with
a convective envelope or a giant, by subscript \( 2 \). Both stars
move on circular orbits with an orbital distance \( A \). We assume
that the secondary corotates with the orbital motion. This is justified
by the short timescales of orbital circulatization and synchronization
for such systems  \citep[e.g.,][]{Zahn77}.
The mass ratio of both stars is defined as\begin{equation}
\label{Eq:Def:q}
q=\frac{M_{2}}{M_{1}}.
\end{equation}
The secondary almost fills its critical Roche lobe and therefore loses
mass through the inner Lagrangian point \( L_{1} \) to its compact
companion.

\subsection{The mass transfer rate}

Since the donor star does not have a sharp rim, the mass loss rate
must be a continuous function of the difference\begin{equation}
\label{Eq:Def:DeltaR}
\Delta R=R_{2}-R_{\text {R,2}}
\end{equation}
 between the radius \( R_{2} \) and the Roche radius \( R_{\text {R,2}} \)
of the star. \kommentar{The effective pressure scale height\begin{equation}
\label{Eq:Hp}
H_{\text {P}}=\frac{kT_{0}R_{\text {R,2}}^{2}}{\mu _{0}\Gamma M_{2}\gamma _{\text {R}}(q)}
\end{equation}
at the photosphere is a measure of how well-defined the rim of the
star is. Here, \( T_{0} \) denotes the effective temperature, \( \mu _{0} \)
the photospheric mean molecular weight, \( k \) the Boltzmann constant,
\( \Gamma  \) the gravitational constant, and \( \gamma _{\text {R}}(q)\approx 1 \)
takes into account the deviation of the gravitational potential from
sphericity  \citep[for more details see][]{Ritter88}.
\( \frac{H_{\text {P}}}{R_{2}} \) is of order \( \sim 10^{-4} \)
for low-mass main sequence (MS) stars and up to \( \sim 10^{-2} \)
for giants. Therefore, the mass loss rate \( \dot{M}_{2} \) is a
rather steep function of \( \Delta R \).} For computing the mass
loss rate we use\begin{equation}
\label{Eq:MassTransferRate}
\dot{M}_{2}=-\dot{M}_{0}\, \exp \left( \frac{\Delta R}{H_{\text {P}}}\right) ,
\end{equation}
where \( \dot{M}_{0}>0 \) is a weakly varying function of the system
parameters \( M_{1} \), \( M_{2} \), \( A \), and the photospheric
values of the donor, and \( H_{\text {P}} \) denotes the photospheric
pressure scale height  \citep[for details see][]{Ritter88}.
Strictly speaking, Eq.~(\ref{Eq:MassTransferRate}) is applicable
only to a donor star which slightly underfills its Roche lobe. This
is the case for most of the systems we are interested in. In any case,
the conclusions of this paper are independent of the exact form of
the mass transfer prescription as long as the characteristic \corr{scale
length \( H \), on which \( \dot{M}_{2} \) increases by a factor
of \( \text {e} \), fulfills}\begin{equation}
\label{Eq:Def:H}
H:=\left( \frac{1}{\dot{M}_{2}}\Ableitung{\dot{M}_{2}}{\Delta R}\right) ^{-1}\ll R_{2}.
\end{equation}
 \kommentar{The length scale \( H \) defines, by which amount \( \Delta R \)
has to change in order to change \( \dot{M}_{2} \) by a factor of
\( \text {e} \).} We note that in (\ref{Eq:MassTransferRate}) the
characteristic scale length \corr{is \( H=H_{\text {P}} \) and} that
\( \frac{H_{\text {P}}}{R_{2}} \) is of order \( \sim 10^{-4} \)
for low-mass MS stars and up to \( \sim 10^{-2} \) for giants.

\subsection{The accretion geometry and the irradiating flux}

We assume that the matter streaming through the \( L_{1} \)-point
is collected in an accretion disk surrounding the primary and that
a fraction \( 0\leq \eta \leq 1 \) of the transferred matter is finally
accreted, i.e.:\begin{equation}
\label{Eq:Def:eta}
\dot{M}_{1}=-\eta \dot{M}_{2}.
\end{equation}
The fraction \( 1-\eta  \) is lost from the system and carries away
angular momentum, which is parametrized in terms of\begin{equation}
\nu =\Ableitung{\ln J}{\ln M},
\end{equation}
where \( J \) is the orbital angular momentum and \( M=M_{1}+M_{2} \)
the total mass of the system. This mode of orbital angular momentum
loss is usually referred to as consequential angular momentum loss
 \citep[CAML, see, e.g.,][]{King95a}.
If the matter is lost with the specific orbital angular momentum of
the primary, then \( \nu =q \).

The accretion process releases gravitational energy and produces accretion
luminosity, mostly on or nearby the compact star so that the source
of the accretion luminosity can be treated as a point source. Hence,
the irradiating flux at a distance \( d \) is given by\begin{equation}
\label{Eq:Def:Firr:1}
F_{\text {irr}}=\frac{\alpha _{\text {accr}}}{4\pi d^{2}}\frac{\Gamma M_{1}}{R_{1}}\dot{M}_{1}.
\end{equation}
\( \Gamma  \) is the gravitational constant, and \( \alpha _{\text {accr}} \)
takes into account that the gravitational energy is not necessarily
completely released as accretion luminosity and that the accretion
luminosity is not necessarily radiated isotropically with respect
to the primary.

On the long term the system can lose additional matter and angular
momentum, e.g., by nova explosions so that the long-term average value
\( \bar{\eta } \) can be less than \( \eta  \), even \( \bar{\eta }<0 \)
is possible  \citep[e.g.,][]{Hameury89, Prialnik95}.
In this case \( \bar{\eta } \) determines the (long-term) mass and
angular momentum balance while \( \eta  \) determines the momentary
accretion rate and therefore the accretion luminosity. As has been
shown by  \citet{Schenker98}, describing the
long-term evolution by an average \( \bar{\eta } \) is reasonable.

\subsection{The reflection effect and the intrinsic flux}

A star that is illuminated by its companion shows a reflection effect
 \citep[for a review see][]{Vaz85}.
Accordingly, the star has an illuminated, brighter and hotter side
with an irradiation-dependend effective temperature \( T_{\text {irr}} \)
and an unilluminated, darker and cooler side with an effective temperature
\( T_{0} \). The intrinsic flux \( F_{\text {int}} \) through a
surface element is defined as the net flux, i.e., the difference between
the incoming flux \( F_{\text {irr}} \) and the outgoing flux \( \sigma T^{4}_{\text {irr}} \),
where \( \sigma  \) denotes the Stephan-Boltzmann constant. It is
convenient to write the irradiating flux in units of the intrinsic
flux \( F_{0}=\sigma T_{0}^{4} \) on the unilluminated side:\begin{equation}
f_{\text {irr}}=\frac{F_{\text {irr}}}{F_{0}}.
\end{equation}

Typically, the reflection effect is discussed in terms of the bolometric
reflection albedo \begin{equation}
\label{Eq:Def:Wbol}
w_{\text {bol}}=\frac{\sigma T_{\text {irr}}^{4}-\sigma T_{0}^{4}}{F_{\text {irr}}}
\end{equation}
but for the purpose of this paper it is more convenient to use the
intrinsic flux in units of \( F_{0} \):\begin{equation}
\label{Eq:Def:Fint}
f_{\text {int}}=\frac{F_{\text {int}}}{F_{0}}=\frac{\sigma T_{\text {irr}}^{4}-F_{\text {irr}}}{F_{0}}=\frac{T^{4}_{\text {irr}}}{T_{0}^{4}}-f_{\text {irr}}.
\end{equation}
For stars in radiative equilibrium the irradiating flux \( f_{\text {irr}} \)
is reprocessed and completely reemitted. They show \( w_{\text {bol}}=1 \)
and \( f_{\text {int}}=1 \) all over the surface. Stars with convective
envelopes are by definition not in radiative equilibrium and for them
\( w_{\text {bol}} \) and \( f_{\text {int}} \) can be less than
unity.

Convection in the interior of stars is efficient, i.e., adiabatic.
It was first stated by  \citet{Rucinski69}
that the entropy, which is constant in the adiabatic convection zone,
must determine the intrinsic flux through the outer layers. This means
that, as was first recognized by  \citet{Nordlund85},
the temperature gradient must be flatter on the illuminated, hotter
side than on the unilluminated, cooler side: Since the temperature
gradient in the adiabatic convection zone is essentially tied to the
adiabatic temperature gradient, the temperature gradient in the subphotospheric,
superadiabatic layers must be flatter on the illuminated side. Therefore,
energy transport through the illuminated outer layers is inhibited.
In fact, the superadiabatic zone acts like a valve for the energy
flow. This can be unterstood in terms of a simple one-zone model \cite{Ritter95, Ritter00}.

\kommentar{While \( f_{\text {int}}\leq 1 \) is an upper limit, a
lower limit\footnote{%
This is not a strict limit because under certain circumstances (small
angle of incidence, depending on atmospheric structure and opacity)
\( w_{\text {bol}} \) can become slightly negative so that the irradiated
star is \emph{cooler} than the unilluminated one \cite{Nordlund85}.
} can be obtained for \( f_{\text {irr}}\leq 1 \) by using \( T_{\text {irr}}\geq T_{0} \)
in (\ref{Eq:Def:Fint}):\begin{equation}
\label{Eq:Fint:LowerLimit}
1-f_{\text {irr}}\leq f_{\text {int}}
\end{equation}
which is equivalent to \( w_{\text {bol}}\geq 0 \). For \( f_{\text {irr}}>1 \)
the lower limit is \( 0\leq f_{\text {int}} \). As can be seen from
(\ref{Eq:Def:Fint}), \( f_{\text {int}} \) decreases more slowly,
than \( f_{\text {irr}} \) increases, since the temperature \( T_{\text {irr}} \)
of an illuminated surface element increases with increasing \( f_{\text {irr}} \):\begin{equation}
\label{Eq:dFintdFirr:limits}
0\leq -\Ableitung{f_{\text {int}}}{f_{\text {irr}}}=1-\Ableitung{}{f_{\text {irr}}}\frac{T_{\text {irr}}^{4}}{T_{0}^{4}}\leq 1.
\end{equation}
For a more detailed discussion about \( f_{\text {int}} \) see
\citet{Hameury97}\footnote{%
Those authors denote \( f_{\text {irr}} \) by \( x \), \( f_{\text {int}} \)
by \( G(x) \) and \( \Ableitung{f_{\text {int}}}{f_{\text {irr}}} \)
by \( -g(x) \).
}.}

 \citet[][ App.~A]{Ritter00}
have pointed out that lateral energy transport within the superadiadiabatic
convection zone is negligible since the radial temperature gradient
in the superadiabatic layers is much larger than the lateral temperature
gradient. Hence, irradiation can be treated as a local effect and
for every surface element separately. \corr{While this is most likely
a good approximation for weak irradiation, i.e., typical CVs, it is
possible that for sufficiently strong irradiation, i.e., typical LMXBs,
non-local effects like circulations, which transport heat from the
illuminated to the unilluminated side, become non-negligible. We will
discuss this topic in more detail in Sec.~\ref{Sec:Numerics:nonlocal}.}

\kommentar{There are indications by numerical computations that this
is also true for strong irradiation as expected in LMXBs \cite{Beer02}.}

By solving the equations of stellar structure for an unirradiated
and an irradiated surface element simultaneously for a fixed entropy
at the bottom of the adiabatic zone it is possible to compute \( f_{\text {int}}\left( f_{\text {irr}}\right)  \)
numerically. This has been done by  \citet{Hameury97}\footnote{%
Those authors denote \( f_{\text {irr}} \) by \( x \), \( f_{\text {int}} \)
by \( G(x) \) and \( \Ableitung{f_{\text {int}}}{f_{\text {irr}}} \)
by \( -g(x) \).
} for grey atmospheres and perpendicular irradiation for a large grid
of effective temperature \( T_{0} \) and surface gravity \( g=\frac{\Gamma M_{2}}{R_{2}^{2}} \).
The results are available in tabular form and we use them for our
numerical calculations.

The assumption of a grey atmosphere and an irradiated Planck spectrum
is not necessarily very accurate, especially for CVs and LMXBs. Various
properties like a different penetration depth of optical light, UV
and X-rays are neglected. A smaller penetration depth means that the
external radiation field is mainly absorbed and reprocessed in the
optically thin\footnote{%
{}``Optically thin{}'' means optically thin for the outgoing radiation
field. The same atmospheric layers can be optically thick for the
incoming radiation field.
} layers of the atmosphere. This reduces the effect of the irradiation
on the photospheric boundary conditions and therefore the intrinsic
flux is less inhibited. This {}``efficiency{}'' of the irradiation
for a non-grey atmosphere compared to a grey atmosphere will be taken
into account formally by a free parameter \( \alpha _{\text {irr}} \).
If the external radiation field has the same spectrum as the irradiated
\corr{star, then} \( \alpha _{\text {irr}} \) is unity by definition.
\corr{In the case of LMXBs hard X-rays, which can penetrate into the
subphotospheric layers, can be the dominant component of the irradiating
spectrum. But since even in the case of a grey atmosphere (i.e., \( \alpha _{\text {irr}}=1 \))
about \( \exp \left( -\frac{2}{3}\right) \approx \frac{1}{2} \) of
the irradiating flux penetrates directly down to the photosphere at
an optical depth of \( \tau =\frac{2}{3} \)  \citep[see e.g.,][ Eq.~29]{Tout89}
and since even hard X-rays, except for the most energetic photons,
do not penetrate sufficiently deep into the outer layers \cite{Hameury96},
we do not expect that \( \alpha _{\text {irr}} \) can become significantly
greater than unity in typical LMXBs. Therefore, in general \( \alpha _{\text {irr}}\lesssim 1 \).}

Models of non-grey atmospheres irradiated by Planck spectra \cite{Nordlund90, Brett93}
and real stellar spectra \cite{Hauschildt01, Hauschildt02, Barman02}
have been computed. However, those computations are much too time-consuming
for generating tables that would be applicable for our purpose, i.e.,
similar to those by  \citet{Hameury97}. Nevertheless,
it is possible to compare models of irradiated non-grey atmospheres
with models of irradiated grey atmospheres to get an estimate for
the irradiation efficiency~\( \alpha _{\text {irr}} \).

\begin{figure}[htb]
\centering
\resizebox{\hsize}{!}{\includegraphics{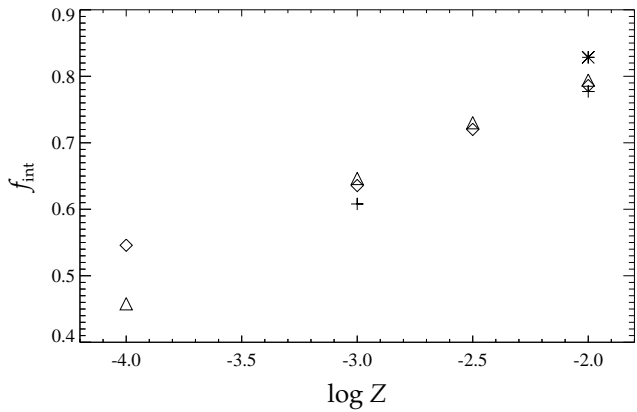}}
\caption{The intrinsic flux \( f_{\text {int}} \) as function of
metallicity \( Z \) for a star with \( T_{0}=4\, 500\, \text {K} \)
and an effective gravity \( \log g=4.5 \) that is irradiated by \( f_{\text {irr}}\approx 1.006 \)
with an angle of incidence \( \beta =47.9\degr  \), i.e., effectively
by \( f_{\text {irr}}\approx 0.674 \). The star is irradiated by
its companion with \( T_{1}=6\, 000\, \text {K} \) and a relative
radius of \( \frac{R_{1}}{A}=\frac{1}{\sqrt{\pi }} \). The data shown
are taken from  \citet{Hameury97} (asterisk)
and  \citet[ table~1]{Nordlund90}: grey atmospheres
(crosses), and non-grey atmospheres irradiated by a Planck spectrum
(triangles) and a real stellar spectrum (diamonds).}
\label{Fig:Nordlund:Metals}
\end{figure}

Since \( f_{\text {int}} \) depends on the penetration depth, irradiation
is more efficient for a smaller opacity and thus for a smaller metallicity.
Fig.~\ref{Fig:Nordlund:Metals} shows \( f_{\text {int}} \) as a
function of metallicity for a specific binary system. Since the source
radiates mainly in the optical, the results for grey atmospheres (crosses)
do not differ much from those for non-grey atmospheres of the irradiated
star, for an irradiated Planck spectrum (triangles) as well as for
a real stellar spectrum (diamonds). The data shown are taken from
 \citet[ table~1]{Nordlund90} and are compared
with results from  \citet{Hameury97} (asterisk).

\begin{figure}[htb]
\centering
\resizebox{\hsize}{!}{\includegraphics{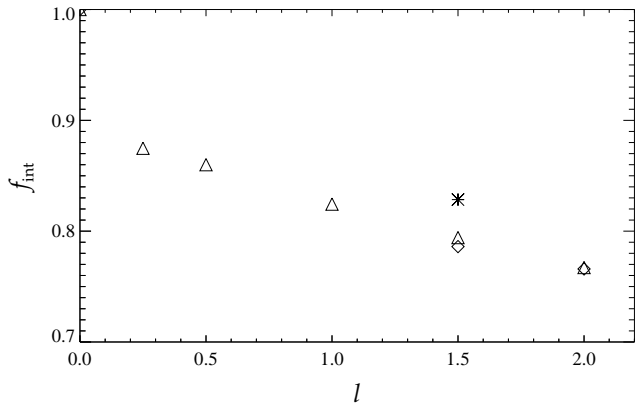}}
\caption{The intrinsic flux \( f_{\text {int}} \) as function of
the mixing length \( l \) in units of the pressure scale height \( H_{\text {P}} \).
Otherwise see Fig.~\ref{Fig:Nordlund:Metals}.}
\label{Fig:Nordlund:Mixing}
\end{figure}

The reduction of \( f_{\text {int}} \) by irradiation also depends
on the model of convection, especially the mixing length \( l \)
as shown in Fig.~\ref{Fig:Nordlund:Mixing}. The data are taken from
 \citet[ table~2]{Nordlund90} and 
\citet{Hameury97}.

\begin{figure}[htb]
\centering
\resizebox{\hsize}{!}{\includegraphics{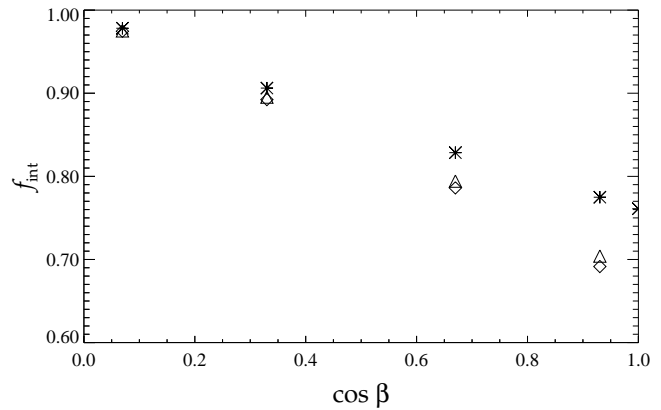}}
\caption{The intrinsic flux \( f_{\text {int}} \) as function of
the angle of incidence \( \beta  \). Otherwise see Fig.~\ref{Fig:Nordlund:Metals}.}
\label{Fig:Nordlund:Angle}
\end{figure}

For a plane-parallel atmosphere the external radiation field as a
function of optical depth \( \tau  \) and angle of incidence \( \beta  \)
is given by\begin{equation}
f_{\text {irr}}(\tau ,\beta )=f_{\text {irr}}(0,0)\, \cos \beta \, \exp \left( -\frac{\tau }{\cos \beta }\right) ,
\end{equation}
 where \( f_{\text {irr}}(0,0) \) is the irradiating flux at \( {\tau =0} \)
\cite{Brett93}. Fig.~\ref{Fig:Nordlund:Angle} shows \( f_{\text {int}} \)
as function of~\( \beta  \). The data from 
\citet[ table~3]{Nordlund90} have been computed for \( f_{\text {irr}}\approx 1.006 \)
and different values of \( \beta  \) while the data from 
\citet{Hameury97} are formally valid only for perpendicular irradiation.
The corresponding data points (asterisks) were computed for perpendicular
irradiation with an {}``effective{}'' flux \( f_{\text {irr}}\cos \beta  \).
Since Fig.~\ref{Fig:Nordlund:Angle} does not show any significant
difference for large angles of incidence \( \left( \cos \beta \rightarrow 0\right)  \),
our treatment of an effective flux \( f_{\text {irr}}\cos \beta  \)
seems to be reasonable.

\begin{figure}[htb]
\centering
\resizebox{\hsize}{!}{\includegraphics{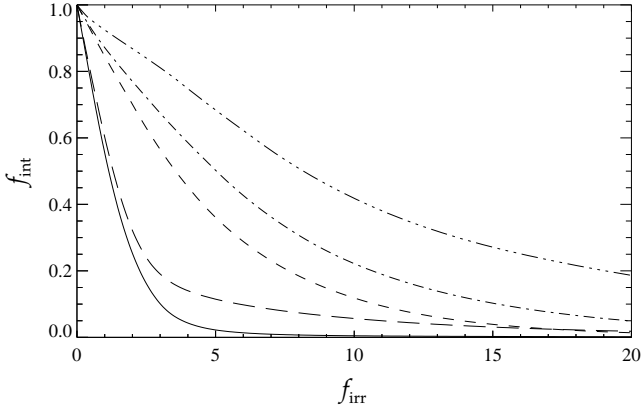}}
\caption{\( f_{\text {int}} \) for models no.~1-4 and~6 of table~\ref{Table:SingleStars}
(solid, long-dashed, short-dashed, dash-dotted and dot-dash-dotted
line, respectively).}
\label{Fig:Fint}
\end{figure}

Since \( f_{\text {irr}} \) can fluctuate on short timescales and
the relation between \( f_{\text {int}} \) and \( f_{\text {irr}} \)
is non-linear, the effect of irradiation does not only depend on the
average \( f_{\text {irr}} \), but also on the amplitude and period
of those fluctuations. This is taken into account formally by another
free parameter \( \alpha _{\text {d}}\leq 1 \) which addresses the
duty cycle of the irradiation. For constant irradiation \( \alpha _{\text {d}} \)
is unity. For time-varying irradiation the efficiency depends on the
(time-averaged) blocked part of the intrinsic flux \( f_{\text {blocked}}:=1-f_{\text {int}} \)
which determines the heating of the convective envelope. Since \( f_{\text {int}} \)
is a convex function for typical stellar envelopes as shown by Fig.~\ref{Fig:Fint},
i.e.,\begin{equation}
f_{\text {int}}\left( a_{1}f_{\text {irr,1}}+a_{2}f_{\text {irr,2}}\right) \leq a_{1}f_{\text {int}}\left( f_{\text {irr,1}}\right) +a_{2}f_{\text {int}}\left( f_{\text {irr,2}}\right) 
\end{equation}
for any \( a_{1} \), \( a_{2}>0 \), \( a_{1}+a_{2}=1 \), which
is equivalent to \( {f_{\text {int}}''\geq 0} \), the efficiency
and therefore \( \alpha _{\text {d}} \) decreases for non-constant
irradiation. Splitting a mean \( f_{\text {irr}} \) into two phases
of \( f_{\text {irr,1}} \) and \( f_{\text {irr,2}} \) increases
the resulting mean \( f_{\text {int}} \) and thus decreases \( f_{\text {blocked}} \).
If the illumination oscillates between an {}``On{}'' state with
high fluxes \( \left( f_{\text {int}}\approx 0\right)  \) and an
{}``Off{}'' state with \( f_{\text {irr}}\approx 0 \), then we
get\begin{equation}
\alpha _{\text {d}}\approx \frac{\tau _{\text {on}}}{\tau _{\text {on}}+\tau _{\text {off}}},
\end{equation}
where \( \tau _{\text {on}} \) and \( \tau _{\text {off}} \) denote
the average duration of the {}``On{}'' and the {}``Off{}'' state,
respectively. We conclude that, e.g., nova-like variables have \( \alpha _{\text {d}}\approx 1 \)
while dwarf novae can have \( \alpha _{\text {d}}\ll 1 \) in case
of long quiescent phases between outbursts.

\begin{table}
\centering

{\footnotesize \begin{tabular}{|c||c|c|c|c|c|c|}
\hline 
Model No.&
1&
2&
3&
4&
5&
6\\
\hline
\hline 
\( \vphantom {\sqrt{0}}X_{\text {c}} \)&
0.71&
0.71&
0.71&
0.05&
0.36&
0\\
\hline 
\( \vphantom {\sqrt{0}}M_{2}/M_{\sun } \)&
0.3&
0.5&
0.8&
0.45&
0.60&
0.8\\
\hline 
\( \vphantom {\sqrt{0}}R_{2}/R_{\sun } \)&
0.284&
0.436&
0.694&
0.695&
0.755&
25.81\\
\hline 
\( \vphantom {\sqrt{0}}\log T_{0}/\text {K} \)&
3.552&
3.590&
3.705&
3.648&
3.738&
3.582\\
\hline 
\( \vphantom {\sqrt{0}}\log L/L_{\sun } \)&
-1.93&
-1.41&
-0.54&
-0.77&
-0.34&
2.10\\
\hline 
\( \vphantom {\sqrt{0}}M_{\text {ce}}/M_{2} \)&
1.0&
0.23&
0.04&
0.17&
0.035&
0.63\\
\hline 
\( \vphantom {\sqrt{0}}10^{4}H_{\text {P}}/R_{2} \)&
0.84&
1.00&
1.50&
2.2&
2.2&
41.1\\
\hline 
\( \vphantom {\sqrt{0}}\zeta _{\text {s}} \)&
-0.31&
-0.08&
1.12&
0.60&
2.4&
0.14\\
\hline 
\( \vphantom {\sqrt{0}}\zeta _{\text {e}} \)&
0.70&
1.04&
0.90&
0.76&
1.45&
-0.2\\
\hline
\end{tabular}}{\footnotesize \par}

\caption{\label{Table:SingleStars}Values of several stellar models. No.~1-3
are unevolved MS stars, 4 is a remnant of thermal timescale mass transfer
of an evolved MS star with an initial mass of \protect\( 1.5\, M_{\sun }\protect \),
an initial central hydrogen abundance of \protect\( 0.06\protect \),
an initially \protect\( 0.6\, M_{\sun }\protect \) white dwarf primary,
\protect\( \bar{\eta }=0.25\protect \), and subsequent mass loss
driven by strong braking according to  \citet{Verbunt-Zwaan}
using \protect\( f_{\text {VZ}}=1\protect \). No.~5 is similar but
with an initial mass of \protect\( 3\, M_{\sun }\protect \), an initial
central hydrogen abundance of \protect\( 0.41\protect \), an initially
\protect\( 1.4\, M_{\sun }\protect \) neutron star primary, and \protect\( \bar{\eta }\protect \)
is determined by an Eddington accretion rate of \protect\( 2\cdot 10^{-8}\, M_{\sun }/\text {yr}\protect \),
otherwise \protect\( \bar{\eta }=1\protect \). No.~6 is a giant.}
\end{table}

In the following we will use the symbol \( f_{\text {irr}} \), respectively
\( F_{\text {irr}} \) for the effective irradiating flux\begin{equation}
\label{Eq:Firr:allg}
f_{\text {irr}}=\frac{F_{\text {irr}}}{F_{0}}=-\frac{\alpha \eta }{4\pi d^{2}\sigma T_{0}^{4}}\frac{\Gamma M_{1}}{R_{1}}\dot{M}_{2}
\end{equation}
instead of (\ref{Eq:Def:Firr:1}), where the total efficiency \( \alpha  \)
is defined as\begin{equation}
\label{Eq:Def:alpha}
\alpha =\alpha _{\text {d}}\alpha _{\text {irr}}\alpha _{\text {accr}}
\end{equation}
 and \( \dot{M}_{1} \) is replaced by (\ref{Eq:Def:eta}).

\subsection{The irradiation geometry}

\subsubsection{The effectively blocked surface fraction}

Using the definition of \( f_{\text {int}} \) in (\ref{Eq:Def:Fint})
the net luminosity of the star is given by\begin{equation}
\label{Eq:Def:Lint}
L_{\text {int}}=\int _{0}^{2\pi }\int _{0}^{\pi }f_{\text {int}}\left( f_{\text {irr}}\right) \, \sigma T_{0}^{4}R_{2}^{2}\, \sin \vartheta \, \text {d}\vartheta \, \text {d}\varphi ,
\end{equation}
but it is more convenient to write \( L_{\text {int}} \) as a function
of \( R_{2} \) and \( T_{0} \) on the unirradiated side and an effectively
blocked fraction of the surface \( s \):\begin{equation}
\label{Eq:StephanBoltzmann}
L_{\text {int}}=4\pi (1-s)R_{2}^{2}\sigma T_{0}^{4}
\end{equation}
with \begin{equation}
\label{Eq:Def:s}
s=\frac{1}{4\pi }\int _{0}^{2\pi }\int _{0}^{\pi }\left[ 1-f_{\text {int}}\left( f_{\text {irr}}\right) \right] \, \sin \vartheta \, \text {d}\vartheta \, \text {d}\varphi .
\end{equation}
(\ref{Eq:StephanBoltzmann}) is a modified Stephan-Boltzmann equation
for an asymmetrically irradiated star. The star is described as if
a fraction \( s \) of its surface were completely blocked by the
irradiation and a fraction \( 1-s \) were unirradiated. Eq.~(\ref{Eq:StephanBoltzmann})
allows us to describe an asymmetrically irradiated star by a \( 1 \)~D
stellar model where the equations of stellar structure are solved
{}``on the unilluminated side{}''.

\begin{figure}[htb]
\centering
\resizebox{\hsize}{!}{\includegraphics{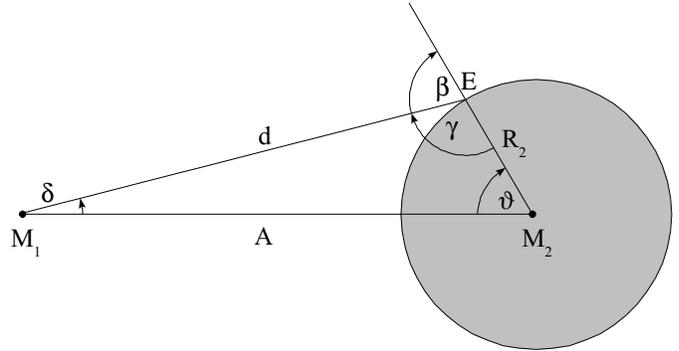}}
\caption{Irradiation geometry. A surface element \( E \) is irradiated
by the source with an angle of incidence \( \beta  \) to the surface
normal.}
\label{Fig:Geometry}
\end{figure}

Due to the axial symmetry of the external radiation field (\ref{Eq:Def:s})
can be simplified:\begin{equation}
\label{Eq:s:PointSource}
s=\frac{1}{2}\int ^{\vartheta _{\text {max}}}_{0}\left[ 1-f_{\text {int}}\left( f_{\text {irr}}\right) \right] \, \sin \vartheta \, \text {d}\vartheta .
\end{equation}
The maximum fraction of the surface that can be blocked by irradiation
is determined by the system geometry. As can be seen from Fig.~\ref{Fig:Geometry},
only surface elements with \( 0\leq \vartheta <\vartheta _{\text {max}} \)
are illuminated while surface elements with \( \vartheta _{\text {max}}\leq \vartheta \leq \pi  \)
can not see the source. \( \vartheta _{\text {max}} \) is defined
by\begin{equation}
\label{Eq:cosThetaMax}
\cos \vartheta _{\text {max}}=\frac{R_{2}}{A}.
\end{equation}
Since \( f_{\text {int}}=1 \) on unirradiated surface elements and
\( f_{\text {int}}\geq 0 \) on irradiated surface elements, (\ref{Eq:s:PointSource})
yields the maximum value of \( s \):\begin{equation}
s\leq \frac{1}{2}\int _{0}^{\vartheta _{\text {max}}}\sin \vartheta \, \text {d}\vartheta =\frac{1}{2}\left( 1-\cos \vartheta _{\text {max}}\right) =:s_{\text {max}}.
\end{equation}
Obviously, we have \( s_{\text {max}}\leq \frac{1}{2} \), and for
typical system parameters it is of order of \( 0.3-0.4 \).

\subsubsection{The constant flux model}

The simplest conceivable irradiation model is the constant flux model
\cite{Ritter95, Ritter00} which assumes that \( f_{\text {irr}} \)
has the constant average value \begin{equation}
\label{Eq:Firr:ConstantFlux}
f_{\text {irr}}=\frac{1}{2}\left\langle f_{\text {irr}}\right\rangle :=-\frac{\alpha \eta }{8\pi A^{2}\sigma T_{0}^{4}}\frac{\Gamma M_{1}}{R_{1}}\dot{M}_{2}
\end{equation}
over the irradiated surface where \( \left\langle f_{\text {irr}}\right\rangle  \)
denotes the irradiating flux at a distance \( A \) according to (\ref{Eq:Firr:allg}).
Then (\ref{Eq:s:PointSource}) yields:\begin{equation}
\label{Eq:s:ConstantFlux}
s=s_{\text {max}}\left[ 1-f_{\text {int}}\left( \left\langle f_{\text {irr}}\right\rangle \right) \right] .
\end{equation}
\kommentar{From this formulation it can be easily seen that even if
a part of the secondary's surface is shadowed by the rim of the accretion
disk the behaviour of of the star does not change fundamentally, but
only \( s_{\text {max}} \) is decreased.}

\subsubsection{The point source model}

A more realistic model is the point source model \cite{King96, Ritter00}
where the irradiation source is treated as a point source as shown
in Fig.~\ref{Fig:Geometry}. In this case the effective flux for
a surface element \( E \) is given by\begin{equation}
\label{Eq:Firr:PointSource}
f_{\text {irr}}=-\frac{\alpha \eta }{4\pi A^{2}\sigma T_{0}^{4}}\frac{\Gamma M_{1}}{R_{1}}\dot{M}_{2}\cos \beta =\left\langle f_{\text {irr}}\right\rangle \cos \beta ,
\end{equation}
where \( \cos \beta  \) can be expressed by\begin{equation}
\cos \beta =\frac{\cos \vartheta -\cos \vartheta _{\text {max}}}{\left( 1-2\cos \vartheta \cos \vartheta _{\text {max}}+\cos ^{2}\vartheta _{\text {max}}\right) ^{\frac{3}{2}}}=:h(\vartheta ).
\end{equation}

\subsection{Thermal relaxation}

At turn-on of mass transfer and thus of irradiation a certain effective
fraction \( s \) of the surface of the donor star is immediately
blocked by irradiation. As a consequence the star is out of thermal
equilibrium because the (unchanged) nuclear energy production \( L_{\text {nuc}}=L_{0} \)
is higher than the (reduced) intrinsic luminosity \( L_{\text {int}}=(1-s)L_{0} \).
The reaction of a star to spontaneous blocking of its intrinsic luminosity
has been discussed by  \citet{Spruit82} in
connection with star spots: On short timescales the changes of the
photospheric values on the unilluminated surface are negligible; on
a thermal timescale of the convective envelope \( \tau _{\text {ce}} \)
the star relaxes thermally according to the modified outer boundary
conditions by adjusting \( R_{2} \), \( T_{0} \), and \( L_{\text {nuc}} \).
 \citet{Ritter94, Ritter00} give an analytical
estimate for the effect of irradiation on the thermal equilibrium
values of low-mass MS stars. The effective temperature remains basically
unchanged and the radius increases by about \( \left( 1-s\right) ^{-0.1} \),
i.e., \( \leq 5\% \) for typical values of \( s \).

For our numerical computations we compute \( \dot{R}_{2} \) by solving
the equations of stellar structure for the donor star. This will be
described in more detail in Sect. \ref{Chap:Code}.

\subsection{Mass transfer cycles}

\label{Sec:InputPhysics:Cycles}

\begin{figure}[htp]
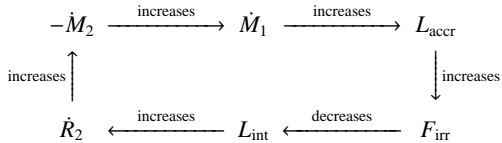

\begin{displaymath}
\begin{CD}
\hspace{1.5cm}\dot{-M_2} @>{\quad{}\text{increases}\quad{}}>>
\dot{M}_{1} @>{\quad{}\text{increases}\quad{}}>> 
L_{\text{accr}}\\
\hspace{1.5cm}@A{\text{increases}}AA 
@. @VV{\text{increases}}V\\
\hspace{1.5cm}\dot{R}_2@<{\quad{}\text{increases}\quad{}}<< 
L_{\text{int}}
@<{\quad{}\text{decreases}\quad{}}<< F_{\text{irr}}
\end{CD}
\end{displaymath}
\caption{Schematic diagram of the irradiation feedback mechanism at
turn on of mass transfer.}
\label{Fig:feedback}
\end{figure}

Fig.~\ref{Fig:feedback} shows schematically how the feedback between
mass transfer rate, irradiation, and thermal relaxation works: As
soon as mass transfer starts, the accretion rate \( \dot{M}_{1} \)
and also the accretion luminosity \( L_{\text {accr}} \) increase.
This enhances the irradiating flux \( F_{\text {irr}} \) which reduces
the intrinsic luminosity \( L_{\text {int}} \). Thus, the thermal
expansion rate \( \dot{R}_{2} \) of the star is increased which in
turn leads to an increasing mass transfer rate \( -\dot{M}_{2} \).

Yet, this feedback can not work forever because thermal relaxation
saturates at last after about a Kelvin-Helmholtz timescale \( \tau _{\text {KH}} \).
But before the star reaches a new equilibrium state, the expansion
rate starts dropping and the mass transfer rate is reduced. Then the
feedback works into the opposite direction: \( \dot{M}_{1} \) and
\( L_{\text {accr}} \) decrease and so does \( F_{\text {irr}} \).
Therefore, \( L_{\text {int}} \) increases and the star starts to
shrink. This reduces the mass transfer even more. In extreme cases
the mass transfer even stops and the system appears as a detached
system until the next cycle starts.

\section{The analytical model}

\label{Chap:AnalyticalModel}

We will now derive an analytical model that describes the onset of
mass transfer cycles. In principle, this has already been done by
 \citet{King96} using a slightly different
formalism. All expressions derived in section \ref{Chap:AnalyticalModel}
and also \ref{Chap:LinearStability} will be applied to the analytical
model only and will not be used for the numerical computations.

\subsection{The governing differential equations}

The time evolution of the Roche radius \( R_{\text {R,2}} \) can
be written as\begin{equation}
\label{Eq:dRR-dt}
\Ableitung{\ln R_{\text {R,2}}}{t}=\zeta _{\text {R}}\Ableitung{\ln M_{2}}{t}+2\left( \Ableitung{\ln J}{t}\right) _{\text {sys}},
\end{equation}
where subscript \( _{\text {sys}} \) denotes the systemic angular
momentum loss at constant total mass \( M \) (e.g., angular momentum
loss by gravitational radiation) and\begin{equation}
\label{Eq:zetaR}
\zeta _{\text {R}}=\left( 1-\bar{\eta }\right) \left( 2\nu +1\right) \frac{q}{1+q}-2\left( 1-\bar{\eta }q\right) +\left( 1+\bar{\eta }q\right) \Ableitung{f_{\text {R}}}{\ln q}
\end{equation}
the mass radius exponent of the critical Roche radius  \citep[e.g.,][]{Ritter99}\footnote{%
In that paper \( q \) is defined as \( \frac{M_{1}}{M_{2}} \), i.e.,
inverse of the definition here.
}. Angular momentum loss by CAML is taken into account by \( \zeta _{\text {R}} \),
and the dimensionless Roche radius \begin{equation}
\label{Eq:Def:fR}
f_{\text {R}}=\frac{R_{\text {R,2}}}{A}
\end{equation}
 is available in tabulated form \cite{Mochnacki}. 

The time evolution of the radius \( R_{2} \) can be splitted into
three terms: The adiabatic reaction at constant entropy and chemical
composition, the thermal relaxation at constant chemical composition,
and the nuclear evolution:\begin{equation}
\label{Eq:dR-dt}
\Ableitung{\ln R_{2}}{t}=\zeta _{\text {s}}\Ableitung{\ln M_{2}}{t}+\left( \Ableitung{\ln R_{2}}{t}\right) _{\text {th}}+\left( \Ableitung{\ln R_{2}}{t}\right) _{\text {nuc}}
\end{equation}
with the adiabatic mass radius exponent \( \zeta _{\text {s}} \). 

The time evolution of the equilibrium radius \( R_{\text {e,2}} \)
of the unilluminated star is given by\begin{equation}
\label{Eq:dRe-dt}
\Ableitung{\ln R_{\text {e,2}}}{t}=\zeta _{\text {e}}\Ableitung{\ln M_{2}}{t}+\left( \Ableitung{\ln R_{\text {e,2}}}{t}\right) _{\text {nuc}},
\end{equation}
where \( \zeta _{\text {e}} \) denotes the thermal equilibrium mass
radius exponent  \citep[e.g.,][]{Webbink85}.

With the timescale of nuclear evolution\begin{equation}
\label{Eq:Def:tau-nuc}
\tau _{\text {nuc}}=\left( \Ableitung{\ln R_{2}}{t}\right) _{\text {nuc}}^{-1}\approx \left( \Ableitung{\ln R_{\text {e,2}}}{t}\right) ^{-1}_{\text {nuc}},
\end{equation}
 the timescale of systemic angular momentum loss\begin{equation}
\label{Eq:Def:tau-J}
\tau _{\text {J}}=-\left( \Ableitung{\ln J}{t}\right) _{\text {sys}}^{-1},
\end{equation}
 and the timescale of thermal relaxation\begin{equation}
\label{Eq:Def:tau-th}
\tau _{\text {th}}=\left( \Ableitung{\ln R_{2}}{t}\right) _{\text {th}}^{-1}
\end{equation}
we define the driving timescale for the mass loss by\begin{equation}
\label{Eq:Def:tau-d}
\frac{1}{\tau _{\text {d}}}=\frac{1}{\tau _{\text {nuc}}}+\frac{2}{\tau _{\text {J}}},
\end{equation}
 and the driving timescale including the thermal relaxation by\begin{equation}
\label{Eq:Def:tau-d-prime}
\frac{1}{\tau '_{\text {d}}}=\frac{1}{\tau _{\text {d}}}+\frac{1}{\tau _{\text {th}}}.
\end{equation}

Using (\ref{Eq:dRR-dt}) and (\ref{Eq:dR-dt}) and taking into account
\( R_{2}\approx R_{\text {R,2}} \) the time evolution of \( \Delta R=R_{2}-R_{\text {R,2}} \)
can be written as\begin{equation}
\label{Eq:dDeltaR-dt}
\Ableitung{}{t}\Delta R=\left( \zeta _{\text {s}}-\zeta _{\text {R}}\right) R_{2}\frac{\dot{M}_{2}}{M_{2}}+\frac{R_{2}}{\tau _{\text {th}}}+\frac{R_{2}}{\tau _{\text {d}}}.
\end{equation}

Using (\ref{Eq:dR-dt}) and (\ref{Eq:dRe-dt}) and assuming\footnote{%
This assumption is not necessarily fulfilled, especially for giants,
but it simplifies the following equations and does not affect the
results significantly.
} \( R_{2}\approx R_{\text {e,2}} \) the time evolution of \begin{equation}
\label{Eq:Def:DeltaRe}
\Delta R_{\text {e}}=R_{2}-R_{\text {e,2}}
\end{equation}
 can be written as\begin{equation}
\label{Eq:dDeltaRe-dt}
\Ableitung{}{t}\Delta R_{\text {e}}=\left( \zeta _{\text {s}}-\zeta _{\text {e}}\right) R_{2}\frac{\dot{M}_{2}}{M_{2}}+\frac{R_{2}}{\tau _{\text {th}}}.
\end{equation}
(\ref{Eq:dDeltaR-dt}) and (\ref{Eq:dDeltaRe-dt}) are the governing
differential equations and are analogous to Eqs. {[}11{]} and {[}12{]}
of  \citet{King96}.

\subsection{The stationary state}

One can easily see from (\ref{Eq:MassTransferRate}) that the mass
transfer rate is stationary if and only if \( \Delta R \) is constant.
Setting the left-hand side of (\ref{Eq:dDeltaR-dt}) to zero yields
the stationary mass transfer rate\begin{equation}
\label{Eq:Mdot:stationary}
\overline{\dot{M}}_{2}=-\frac{M_{2}}{\zeta _{\text {s}}-\zeta _{\text {R}}}\left( \frac{1}{\tau _{\text {th}}}+\frac{1}{\tau _{\text {d}}}\right) =-\frac{1}{\zeta _{\text {s}}-\zeta _{\text {R}}}\frac{M_{2}}{\tau '_{\text {d}}}.
\end{equation}
Additionally, the thermal relaxation\begin{equation}
\label{Eq:Def:K}
K:=\frac{R_{2}}{\tau _{\text {th}}}
\end{equation}
is stationary if \( \Delta R_{\text {e}} \) is also constant. Setting
the left-hand side of (\ref{Eq:dDeltaRe-dt}) to zero and inserting
(\ref{Eq:Mdot:stationary}) yields for the stationary thermal relaxation
\( \overline{K} \):\begin{equation}
\label{Eq:K:stationary}
\frac{\overline{K}}{R_{2}}=\frac{1}{\, \overline{\tau }_{\text {th}}}=\frac{\zeta _{\text {s}}-\zeta _{\text {e}}}{\zeta _{\text {e}}-\zeta _{\text {R}}}\frac{1}{\tau _{\text {d}}}.
\end{equation}
From (\ref{Eq:Mdot:stationary}) and (\ref{Eq:K:stationary}) it can
be seen immediatly that the stationary solution is unique if there
is any at all because \( \overline{\dot{M}}_{2} \) and \( \overline{K} \)
depend on the system parameters \( \zeta _{\text {s}} \), \( \zeta _{\text {e}} \),
\( \zeta _{\text {R}} \), and \( \tau _{\text {d}} \) only. By inserting
(\ref{Eq:K:stationary}) into (\ref{Eq:Mdot:stationary}) the timescale
of mass transfer can be expressed by the driving timescale:\begin{equation}
\label{Eq:Relation:tauM_taud}
\tau _{\text {M}}:=-\frac{M_{2}}{\, \overline{\dot{M}}_{2}}=\left( \zeta _{\text {e}}-\zeta _{\text {R}}\right) \tau _{\text {d}}.
\end{equation}

\section{Linear stability analysis}

\label{Chap:LinearStability}

To determine the conditions for stability of the stationary solution
we perform a linear stability analysis  \citep[see, e.g.,][]{Guckenheimer}
on the vector field\begin{equation}
\vec{F}\left( \begin{array}{c}
\Delta R\\
\Delta R_{\text {e}}
\end{array}\right) =\left( \begin{array}{c}
\left( \zeta _{\text {s}}-\zeta _{\text {R}}\right) R_{2}\frac{\dot{M}_{2}}{M_{2}}+K+\frac{R_{2}}{\tau _{\text {d}}}\\
\left( \zeta _{\text {s}}-\zeta _{\text {e}}\right) R_{2}\frac{\dot{M}_{2}}{M_{2}}+K
\end{array}\right) 
\end{equation}
around the stationary solution \( \left( \overline{\Delta R},\overline{\Delta R}_{\text {e}}\right)  \),
which is a fixed point of \( \vec{F} \), i.e., \( \vec{F}\left( \overline{\Delta R},\overline{\Delta R}_{\text {e}}\right) =\vec{0} \).
\kommentar{For this analysis it is irrelevant whether the system parameters
\( \zeta _{\text {s}} \), \( \zeta _{\text {e}} \), \( \zeta _{\text {R}} \)
and \( M_{2} \) evolve with time as long as they do not depend on
\( \Delta R \) and \( \Delta R_{\text {e}} \).}

\subsection{The necessary criterion}

According to the theorem of Hartmann-Grobmann a fixed point \( \vec{x} \)
of a vector field \( \vec{F} \) is stable, if all eigenvalues of
the Jacobi matrix \( \vec{D}\vec{F} \) of \( \vec{F} \) at \( \vec{x} \)
have a negative real part, and \( \vec{x} \) is unstable if at least
one eigenvalue has a positive real part.

Using (\ref{Eq:MassTransferRate}) and neglecting the derivatives
of \( \tau _{\text {nuc}} \) and \( R_{2} \), which is reasonable
for \( \tau _{\text {nuc}} \), \( \tau _{\text {M}}\gg \tau _{\text {KH}} \),
and neglecting also the derivatives of \( \tau _{\text {J}} \) we
get\begin{equation}
\label{Eq:DF}
\vec{D}\vec{F}\left( \begin{array}{c}
\overline{\Delta R}\\
\overline{\Delta R}_{\text {e}}
\end{array}\right) =\left( \begin{array}{cc}
\left( \zeta _{\text {s}}-\zeta _{\text {R}}\right) \frac{R_{2}}{H_{\text {P}}}\frac{\overline{\dot{M}}_{2}}{M_{2}}+\Partiell{K}{\Delta R} & \Partiell{K}{\Delta R_{\text {e}}}\\
\left( \zeta _{\text {s}}-\zeta _{\text {e}}\right) \frac{R_{2}}{H_{\text {P}}}\frac{\overline{\dot{M}}_{2}}{M_{2}}+\Partiell{K}{\Delta R} & \Partiell{K}{\Delta R_{\text {e}}}
\end{array}\right) 
\end{equation}
at the fixed point. 

\( \vec{D}\vec{F} \) has the eigenvalues\begin{equation}
\label{Eq:DF:eigenvalues}
\lambda _{1,2}=\frac{1}{2}\tr \vec{D}\vec{F}\pm \sqrt{\frac{1}{4}\tr ^{2}\vec{D}\vec{F}-\det \vec{D}\vec{F}}.
\end{equation}
Since \( \vec{D}\vec{F} \) is a real matrix, the necessary condition
for \( {\Re \lambda _{1,2}<0} \) is\begin{equation}
\label{Eq:criterion:necessary}
\tr \vec{D}\vec{F}=\left( \zeta _{\text {s}}-\zeta _{\text {R}}\right) \frac{R_{2}}{H_{\text {P}}}\overline{\dot{M}}_{2}+\Partiell{K}{\Delta R}+\Partiell{K}{\Delta R_{\text {e}}}<0.
\end{equation}
If the thermal relaxation \( K \) is neglected, Eq. (\ref{Eq:criterion:necessary})
is equivalent to\begin{equation}
\label{Eq:criterion:dynamical}
\zeta _{\text {s}}-\zeta _{\text {R}}>0,
\end{equation}
the well-known criterion for dynamical stability  \citep[e.g.,][]{Webbink85}.
Since only dynamically stable systems can undergo long phases of mass
transfer, we consider only such systems.

The first term in (\ref{Eq:criterion:necessary}) reflects the adiabatic
reaction of the star and its Roche radius on mass loss. For dynamically
stable systems this term is negative and therefore it contributes
to the stabilization of the fixed point. The third term in (\ref{Eq:criterion:necessary})
tells how thermal relaxation changes with the radius of the star compared
to its equilibrium radius. A negative value means that the larger
(smaller) \( R_{2} \), the stronger the thermal relaxation that decreases
(increases) the radius to its (irradiation-dependend) equilibrium
value. Therefore,\begin{equation}
\Partiell{K}{\Delta R_{\text {e}}}<0
\end{equation}
 is a reasonable assumption about the thermal relaxation \( K \).
Thus, the necessary criterion for stability (\ref{Eq:criterion:necessary})
can only be violated if\begin{equation}
\Partiell{K}{\Delta R}>0.
\end{equation}
This means that an increasing (decreasing) \( \Delta R \) and therefore
an increasing (decreasing) mass transfer rate tends to enhance thermal
relaxation which drives the expansion (contraction) of the star even
more. Exactly this kind of feedback is caused by irradiation. The
question is now whether this feedback can be strong enough to overwhelm
the stabilizing terms in (\ref{Eq:criterion:necessary}).

\subsection{The sufficient criterion}

The sufficient criterion for \( \Re \lambda _{1,2}<0 \) is that the
absolute value of the real part of the square root in (\ref{Eq:DF:eigenvalues})
is less than the absolute value of \( \frac{1}{2}\tr \vec{D}\vec{F} \),
otherwise at least one eigenvalue has a positive real part. Using
(\ref{Eq:DF}) it can be shown that this is equivalent to \( \det \vec{D}\vec{F}>0 \)
which is equivalent to\begin{equation}
\label{Eq:criterion:thermal}
\zeta _{\text {e}}-\zeta _{\text {R}}>0.
\end{equation}
This is the well-known criterion for thermal stability  \citep[e.g.,][]{Webbink85}.
Only thermally stable systems can have stable stationary solutions
since thermally unstable systems transfer mass on a thermal timescale.
As a consequence, the thermal relaxation \( K \) can never reach
a stationary value. Therefore, we consider only systems that are thermally
stable.

\subsection{The onset of instability}

There are different possibilities for the dynamics of the system nearby
the fixed point depending on \( \tr \vec{D}\vec{F} \) and \( \det \vec{D}\vec{F} \)
if (\ref{Eq:criterion:dynamical}) and (\ref{Eq:criterion:thermal})
are fulfilled:

\begin{itemize}
\item \( \tr \vec{D}\vec{F}\leq -2\sqrt{\det \vec{D}\vec{F}} \): We have
\( \tr ^{2}\vec{D}\vec{F}\geq 4\det \vec{D}\vec{F} \) so that the
eigenvalues \( \lambda _{1,2} \) in (\ref{Eq:DF:eigenvalues}) are
real and negative. All solutions near the fixed point converge directly
to the fixed point without any oscillation.
\item \( -2\sqrt{\det \vec{D}\vec{F}}<\tr \vec{D}\vec{F}<0 \): We have
\( \tr ^{2}\vec{D}\vec{F}<4\det \vec{D}\vec{F} \) so that the square
root in (\ref{Eq:DF:eigenvalues}) is imaginary. The eigenvalues can
be written as \( \lambda _{1,2}=-\lambda \pm \text {i}\omega  \)
with \( \lambda ,\omega >0 \). All solutions near the fixed point
spiral in.
\item \( \tr \vec{D}\vec{F}=0 \): The eigenvalues \( \lambda _{1,2}=\pm \text {i}\omega  \)
are imaginary. The system undergoes a Hopf bifurcation, when the fixed
point, that is stable for \( \tr \vec{D}\vec{F}<0 \), becomes unstable,
and a stable limit circle arises for \( \tr \vec{D}\vec{F}>0 \).
In principle, this could be proved mathematically in terms of normal
form theory \cite[~theorem 3.4.2 and eq. 3.4.11]{Guckenheimer} but
it requires explicit knowledge of \( \dot{M}_{2}\left( \Delta R\right)  \)
and \( K\left( \Delta R,\Delta R_{\text {e}}\right)  \).
\item \( 0<\tr \vec{D}\vec{F}<2\sqrt{\det \vec{D}\vec{F}} \): We have \( \tr ^{2}\vec{D}\vec{F}<4\det \vec{D}\vec{F} \)
so that the square root in (\ref{Eq:DF:eigenvalues}) is imaginary.
The eigenvalues can be written as \( \lambda _{1,2}=+\lambda \pm \text {i}\omega  \)
with \( \lambda ,\omega >0 \). All solutions near the fixed point
spiral out and converge to the limit cycle sketched in Sect.~\ref{Sec:InputPhysics:Cycles}.
\item \( 2\sqrt{\det \vec{D}\vec{F}}\leq \tr \vec{D}\vec{F} \): We have
\( \tr ^{2}\vec{D}\vec{F}\geq 4\det \vec{D}\vec{F} \) so that the
eigenvalues \( \lambda _{1,2} \) in (\ref{Eq:DF:eigenvalues}) are
real and positive. All solutions near the fixed point diverge directly
without any oscillation and converge to the limit cycle.
\end{itemize}

\subsection{The irradiation feedback}

\begin{figure}[htb]
\centering
\resizebox{\hsize}{!}{\includegraphics{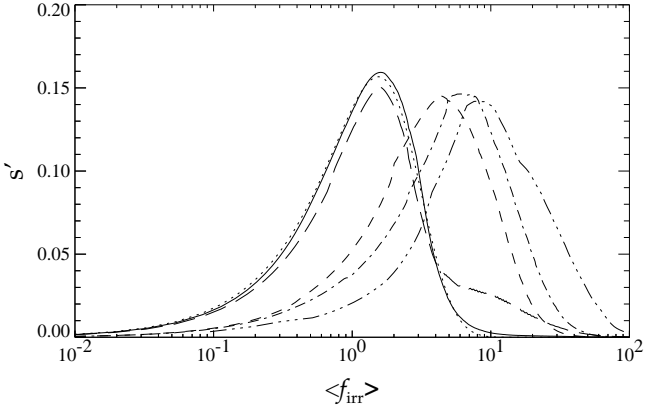}}
\caption{\( s'_{\text {cf}} \) (Eq. (\ref{Eq:s-prime:cf})) for stellar
models no.~1-4 and~6 of table~\ref{Table:SingleStars} (solid,
long-dashed, short-dashed, dash-dotted and dot-dash-dotted line, respectively)
and also \( s'_{k} \) for \( k=1 \) (Eq.~(\ref{Eq:s-prime:k}),
dotted line).}
\label{Fig:s-prime:cf}
\end{figure}

\begin{figure}[htb]
\centering
\resizebox{\hsize}{!}{\includegraphics{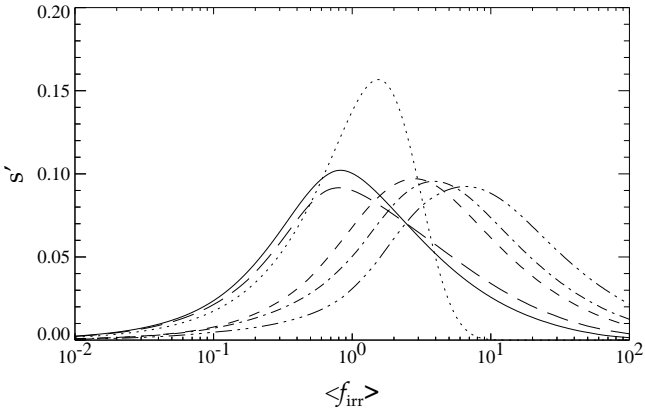}}
\caption{\( s'_{\text {ps}} \) (Eq. (\ref{Eq:s-prime:ps})) for the
same stellar models as in Fig.~\ref{Fig:s-prime:cf}. For comparison
of the point source with the constant flux model also \( s'_{k} \)
for \( k=1 \) (Eq. (\ref{Eq:s-prime:k})) is shown.}
\label{Fig:s-prime:ps}
\end{figure}

Since \( K \) depends on \( \Delta R \) via the intrinsic luminosity
\( L_{\text {int}} \), the derivative \( \Partiell{K}{\Delta R} \)
can be written as\begin{equation}
\label{Eq:dKdx:1}
\Partiell{K}{\Delta R}=\left( \Ableitung{K}{L_{\text {int}}}\right) _{\Delta R_{\text {e}}}\left( \Ableitung{L_{\text {int}}}{\Delta R}\right) _{\Delta R_{\text {e}}}.
\end{equation}
\( L_{\text {int}} \) does not depend on \( \Delta R \) directly
but indirectly via \( s \). Additionally, \( R_{2} \) and \( T_{0} \)
do not depend on \( \dot{M}_{2} \). Therefore we can write:\begin{equation}
\left( \Ableitung{L_{\text {int}}}{\Delta R}\right) _{\Delta R_{\text {e}}}=\left( \Ableitung{L_{\text {int}}}{s}\right) _{R_{2},T_{0}}\Ableitung{s}{\dot{M}_{2}}\Ableitung{\dot{M}_{2}}{\Delta R}.
\end{equation}
From (\ref{Eq:StephanBoltzmann}) we obtain\begin{equation}
\left( \Ableitung{L_{\text {int}}}{s}\right) _{R_{2},T_{0}}=-4\pi \sigma T_{0}^{4}R_{2}^{2}
\end{equation}
and from (\ref{Eq:MassTransferRate})\begin{equation}
\Ableitung{\dot{M}_{2}}{\Delta R}=\frac{\dot{M}_{2}}{H_{\text {P}}}.
\end{equation}
With the definition of\begin{equation}
\label{Def:s-prime}
s':=\dot{M}_{2}\Ableitung{s}{\dot{M}_{2}}=\Ableitung{s}{\ln \left( -\dot{M}_{2}\right) }>0,
\end{equation}
which is a measure for how strongly the blocking of the intrinsic
luminosity changes with \( \dot{M}_{2} \) (or \( f_{\text {irr}} \)),
Eqs. (\ref{Eq:dKdx:1})-(\ref{Def:s-prime}) give\begin{equation}
\label{Eq:dK-dx:feedback}
\Partiell{K}{\Delta R}=-\frac{4\pi \sigma T_{0}^{4}R_{2}^{2}}{H_{\text {P}}}\left( \Ableitung{K}{L_{\text {int}}}\right) _{\Delta R_{\text {e}}}s'.
\end{equation}
Using (\ref{Eq:Def:eta}) and (\ref{Eq:Firr:PointSource}) in (\ref{Eq:s:PointSource})
leads to\begin{equation}
\label{Eq:s-prime:ps}
s'_{\text {ps}}=-\frac{1}{2}\int ^{\vartheta _{\text {max}}}_{0}\Ableitung{f_{\text {int}}}{f_{\text {irr}}}\left\langle f_{\text {irr}}\right\rangle h\left( \vartheta \right) \, \sin \vartheta \, \text {d}\vartheta 
\end{equation}
for the point source model while (\ref{Eq:s:ConstantFlux}) leads
to\begin{equation}
\label{Eq:s-prime:cf}
s'_{\text {cf}}=-\frac{1}{2}s_{\text {max}}\Ableitung{f_{\text {int}}}{f_{\text {irr}}}\left\langle f_{\text {irr}}\right\rangle 
\end{equation}
for the constant flux model. Figs.~\ref{Fig:s-prime:cf} and \ref{Fig:s-prime:ps}
show \( s'_{\text {cf}} \) and \( s'_{\text {ps}} \) for different
stellar models and also for a simple prescription for \( s \) that
was used by  \citet{King95b, King97}:\begin{equation}
s_{k}=s_{\text {max}}\tanh \left( k\left\langle f_{\text {irr}}\right\rangle \right) .
\end{equation}
\( k \) is a free parameter that defines the position of the maximum
of\begin{equation}
\label{Eq:s-prime:k}
s'_{k}=s_{\text {max}}\frac{k\left\langle f_{\text {irr}}\right\rangle }{\cosh ^{2}\left( k\left\langle f_{\text {irr}}\right\rangle \right) }.
\end{equation}
As can be easily seen from (\ref{Eq:s-prime:cf}) and also from (\ref{Eq:s-prime:ps})
or (\ref{Eq:s-prime:k}), \( s' \) vanishes for \( \left\langle f_{\text {irr}}\right\rangle \rightarrow 0 \)
(no irradiation means no feedback) and for \( \left\langle f_{\text {irr}}\right\rangle \rightarrow \infty  \)
(strong irradiation means saturation, i.e., \( s=s_{\text {max}} \)
is constant).

\subsection{The bipolytrope model}

In terms of a bipolytrope model the thermal relaxation \( K \) can
be expressed explicitely \cite[ eq.~32]{Kolb92}. If the star is either
fully convective or the nuclear luminosity of the radiative core is
identical to the total luminosity of the core, which is the case in
the stationary (i.e., quasi-equilibrium) state, then the thermal relaxation
can be expressed as\begin{equation}
\label{Eq:bipolytrope:K}
\frac{K}{R_{2}}=\left( \Ableitung{\ln R_{2}}{t}\right) _{\text {th}}=-{\cal F}\frac{R_{2}\left( L_{\text {int}}-L_{\text {nuc}}\right) }{\Gamma M^{2}_{2}}.
\end{equation}
\( {\cal F} \) is given by Eq. {[}47{]} of 
\citet{King96} for chemically homogeneous stars and can be approximated
by\begin{equation}
\label{Eq:bipolytrope:F}
{\cal F}\approx \frac{7}{3}\frac{M_{2}}{M_{\text {ce}}},
\end{equation}
where \( M_{\text {ce}} \) denotes the mass of the convective envelope
\cite{King96, Ritter00}.

The only quantity in (\ref{Eq:bipolytrope:K}) which depends on \( L_{\text {int}} \)
is \( L_{\text {int}} \) itself:\begin{equation}
\left( \Ableitung{K}{L_{\text {int}}}\right) _{\Delta R_{\text {e}}}=R_{2}\left( \Ableitung{}{L_{\text {int}}}\frac{K}{R_{2}}\right) =-\frac{{\cal F}R^{2}_{2}}{\Gamma M^{2}_{2}}.
\end{equation}
Inserting this into (\ref{Eq:dK-dx:feedback}) gives\begin{equation}
\label{Eq:dK-dx:bipolytrope}
\Partiell{K}{\Delta R}={\cal F}\frac{4\pi \sigma T_{0}^{4}R_{2}^{4}}{H_{\text {P}}\Gamma M^{2}_{2}}s'.
\end{equation}
This can be expressed by quantities of the unilluminated star: In
terms of the Kelvin-Helmholtz timescale\begin{equation}
\label{Eq:Def:KelvinHelmholtz}
\tau _{\text {KH}}=\frac{\Gamma M_{2}^{2}}{R_{\text {e,2}}L_{0}}
\end{equation}
the corresponding timescale of the convective envelope is given by
\begin{equation}
\label{Eq:Def:tau_ce}
\tau _{\text {ce}}=\frac{\tau _{\text {KH}}}{{\cal F}}\sim \frac{3}{7}\frac{M_{\text {ce}}}{M_{2}}\tau _{\text {KH}}.
\end{equation}
Using (\ref{Eq:StephanBoltzmann}) and (\ref{Eq:Def:tau_ce}) Eq.
(\ref{Eq:dK-dx:bipolytrope}) can be written as\begin{equation}
\label{Eq:dK-dx:1}
\Partiell{K}{\Delta R}=\frac{R_{2}}{H_{\text {P}}}\frac{s'}{\tau _{\text {ce}}}\left( \frac{R_{s}}{R_{e}}\right) ^{3}
\end{equation}
and is analogous to Eq.~{[}44{]}, respectively {[}59{]} of 
\citet{King96}. Alternatively, (\ref{Eq:dK-dx:bipolytrope}) can also
be expressed by quantities of the illuminated star:\begin{equation}
\label{Eq:dK-dx:2}
\Partiell{K}{\Delta R}=\frac{R_{2}}{H_{\text {P}}}\frac{s'}{1-s}\frac{1}{\tau _{\text {ce,s}}}.
\end{equation}
In this case the (irradiation-dependent) timescale of the convective
envelope\begin{equation}
\label{Eq:Def:tau:s}
\tau _{\text {ce,s}}=\frac{\tau _{\text {KH,s}}}{{\cal F}}
\end{equation}
is defined via the Kelvin-Helmholtz timescale\begin{equation}
\label{Eq:Def:KelvinHelmholtz:s}
\tau _{\text {KH,s}}=\frac{\Gamma M_{2}^{2}}{R_{2}L_{\text {int}}}
\end{equation}
of the illuminated star taking into account that it can radiate only
through its unilluminated surface with \( L_{\text {int}} \) given
by (\ref{Eq:StephanBoltzmann}).

\subsection{The homology model}

The energy generation rate from hydrogen burning can be approximated
by a simple power law of density~\( \rho  \) and temperature~\( T \):\begin{equation}
\varepsilon _{\text {nuc}}\sim \rho T^{n}.
\end{equation}
For low-mass MS stars, which are considered here, the pp-chain gives
\( \text {n}=5-6 \)  \citep[see, e.g.,][]{Kippenhahn}.
By using homology relations we obtain\begin{equation}
\label{Eq:homology:Lnuc}
L_{\text {nuc}}=L_{0}\left( \frac{R_{2}}{R_{\text {e,2}}}\right) ^{-\left( n+3\right) }
\end{equation}
 for main sequence stars. Since for giants the nuclear luminosity
does not depend on the stellar radius, the following expressions also
apply to giants if we set \( n=-3 \). 

Inserting (\ref{Eq:StephanBoltzmann}) and (\ref{Eq:homology:Lnuc})
into (\ref{Eq:bipolytrope:K}) gives:\begin{equation}
K=-\frac{{\cal F}}{\Gamma M^{2}_{2}}\left[ 4\pi \left( 1-s\right) T_{0}^{4}R_{2}^{4}-L_{0}R^{2}_{\text {e,2}}\left( \frac{R_{2}}{R_{\text {e,2}}}\right) ^{-\left( n+1\right) }\right] .
\end{equation}
The equilibrium radius \( R_{\text {e,2}} \) of the unilluminated
star neither depends on \( \Delta R \) nor on \( \Delta R_{\text {e}} \).
This means\begin{equation}
\Partiell{R_{2}}{\Delta R_{\text {e}}}=\Partiell{}{\Delta R_{\text {e}}}\left( R_{\text {e,2}}+\Delta R_{\text {e}}\right) =1.
\end{equation}
Therefore, the derivative of \( K \) with respect to \( \Delta R_{\text {e}} \)
yields:\begin{equation}
\label{Eq:dK-dy:homology}
\Partiell{K}{\Delta R_{\text {e}}}=-\frac{{\cal F}L_{0}R_{\text {e,2}}\left[ 4\left( 1-s\right) \left( \frac{R_{2}}{R_{\text {e,2}}}\right) ^{3}+\left( n+1\right) \left( \frac{R_{2}}{R_{\text {e,2}}}\right) ^{-\left( n+2\right) }\right] }{\Gamma M_{2}^{2}}.
\end{equation}
This can be written in quantities of the unilluminated star:\begin{equation}
\label{Eq:dK-dy:1}
\Partiell{K}{\Delta R_{\text {e}}}=-\frac{\left[ 4\left( 1-s\right) \left( \frac{R_{2}}{R_{\text {e,2}}}\right) ^{3}+\left( n+1\right) \left( \frac{R_{2}}{R_{\text {e,2}}}\right) ^{-\left( n+2\right) }\right] }{\tau _{\text {ce}}}.
\end{equation}
This result is analogous to Eq.~{[}43{]} of 
\citet{King96} who have computed the derivative\footnote{%
In that paper \( K \) is defined as \( \left( \Ableitung{\ln R}{t}\right) _{\text {th}}^{-1} \).
} of \( \frac{K}{R_{2}} \) while we use the derivative of \( K \).
This leads to different values for the powers of \( \frac{R_{2}}{R_{\text {e,2}}} \)
in their Eq. {[}43{]} compared to our Eq. (\ref{Eq:dK-dy:1}): \( 3 \)
and \( n+2 \) versus \( 4 \) and \( n+1 \). The origin of this
difference is the approximate transition from linear to logarithmic
derivatives in Eq. {[}2{]} of  \citet{King96}.

Alternatively, (\ref{Eq:dK-dy:homology}) can also be expressed in
quantities of the illuminated star. Using (\ref{Eq:Def:tau:s}), (\ref{Eq:Def:KelvinHelmholtz:s}),
(\ref{Eq:homology:Lnuc}) and \( L_{\text {int}}=L_{\text {nuc}} \)
in the (irradiation dependend) equilibrium state we get the more compact
expression:\begin{equation}
\label{Eq:dK-dy:2}
\Partiell{K}{\Delta R_{\text {e}}}=-\frac{n+5}{\tau _{\text {ce,s}}}.
\end{equation}
Finally, we can insert (\ref{Eq:dK-dx:2}) and (\ref{Eq:dK-dy:2})
into the necessary criterion for stability (\ref{Eq:criterion:necessary})
and get\begin{equation}
\label{Eq:criterion:Final:2}
-\Ableitung{\ln \left( 1-s\right) }{\ln \left( -\dot{M}\right) }=\frac{s'}{1-s}<\frac{\tau _{\text {ce,s}}}{\tau '_{\text {d}}}+\frac{H_{\text {P}}}{R_{2}}\left( n+5\right) 
\end{equation}
in quantities of the illuminated star. By inserting (\ref{Eq:dK-dx:1})
and (\ref{Eq:dK-dy:1}) into (\ref{Eq:criterion:necessary}) we get
the equivalent result in quantities of the unilluminated star:\begin{equation}
\label{Eq:criterion:Final:1}
s'<\frac{\tau _{\text {ce}}}{\tau '_{\text {d}}}+\frac{H_{\text {P}}}{R_{2}}\delta ,
\end{equation}
where \( \delta  \) is given by\begin{equation}
\delta =4\left( 1-s\right) \left( \frac{R_{2}}{R_{\text {e,2}}}\right) ^{3}+\left( n+1\right) \left( \frac{R_{2}}{R_{\text {e,2}}}\right) ^{-\left( n+3\right) }\sim n+5.
\end{equation}
If \( s' \) is less than the ratio of \( \tau _{\text {ce}} \) and
\( \tau '_{\text {d}} \) plus a term of the order of several \( \frac{H_{\text {P}}}{R_{2}} \),
then the stationary mass transfer rate is stable. Otherwise it is
unstable and mass transfer cycles occur.

\subsection{Predictions of the analytical model}

Taking (\ref{Eq:Firr:ConstantFlux}), eliminating \( A \) by (\ref{Eq:Def:fR}),
\( \overline{\dot{M}}_{2} \) by (\ref{Eq:Mdot:stationary}) and using
the driving timescale \( \tau '_{\text {d}} \) including the thermal
relaxation from (\ref{Eq:Def:tau-d-prime}) we obtain\begin{equation}
\frac{1}{\tau '_{\text {d}}}=\frac{\zeta _{\text {s}}-\zeta _{\text {R}}}{\alpha \eta }4\pi \sigma T_{0}^{4}\frac{R_{\text {R,2}}^{2}R_{1}}{f_{\text {R}}^{2}\Gamma M_{1}M_{2}}\left\langle f_{\text {irr}}\right\rangle .
\end{equation}
With (\ref{Eq:Def:tau_ce}) the ratio of \( \tau _{\text {ce}} \)
and \( \tau '_{\text {d}} \) can be written as\begin{equation}
\label{Eq:ratio:timescales:analytic}
\frac{\tau _{\text {ce}}}{\tau '_{\text {d}}}=\frac{\zeta _{\text {s}}-\zeta _{\text {R}}}{\alpha \eta {\cal F}}\frac{M_{2}R_{1}R_{\text {R,2}}^{2}}{M_{1}R_{2}^{3}f_{\text {R}}^{2}}\left\langle f_{\text {irr}}\right\rangle .
\end{equation}
This result can be approximated by using (\ref{Eq:bipolytrope:F}),
\( R_{2}\approx R_{\text {R,2}} \) and the approximation \cite{Paczynski71}\begin{equation}
f_{\text {R}}\left( q\right) \approx \left( \frac{8}{3^{4}}\right) ^{\frac{1}{3}}\left( \frac{q}{q+1}\right) ^{\frac{1}{3}}
\end{equation}
 as\begin{equation}
\label{Eq:ratio:timescales:numeric}
\frac{\tau _{\text {ce}}}{\tau '_{\text {d}}}\approx \frac{3}{7}\left( \frac{3^{4}}{8}\right) ^{\frac{2}{3}}\frac{\zeta _{\text {s}}-\zeta _{\text {R}}}{\alpha \eta }\left( 1+q\right) ^{\frac{2}{3}}q^{\frac{1}{3}}\frac{M_{\text {ce}}}{M_{2}}\frac{R_{1}}{R_{2}}\left\langle f_{\text {irr}}\right\rangle .
\end{equation}
This is a function of \( \left\langle f_{\text {irr}}\right\rangle  \)
and its slope basically depends only on \( \alpha \eta  \), the relative
mass of the convective envelope \( \frac{M_{\text {ce}}}{M_{2}} \)
and the ratio of the radii of both stars \( \frac{R_{1}}{R_{2}} \).
The other factors are of order unity for typical system parameters.

For the moment we neglect the last term of order of several \( \frac{H_{\text {P}}}{R_{2}} \)
in criterion (\ref{Eq:criterion:Final:1}) as 
\citet{King96} did. To get mass transfer cycles the blocking of the
intrinsic flux (\( s' \)) as a reaction upon an enhanced mass transfer
rate must increase faster than the convective envelope can relax thermally
(\( \tau _{\text {ce}} \)) compared to the driving time scale of
the mass loss (\( \tau '_{\text {d}} \)). Therefore, systems are
the more susceptible to the onset of mass transfer cycles, the smaller
the timescale of the convective envelope \( \tau _{\text {ce}} \)
and the larger the driving timescale \( \tau _{\text {d}}\sim \tau '_{\text {d}} \)
is.

Accordingly, main sequence stars with thin outer convection zones
\( \left( M_{2}\sim 1\, M_{\sun }\right)  \) have been supposed to
be most susceptible \cite{King96, Ritter00}. Also systems driven
by the longest possible timescale, i.e., the timescale of angular
momentum loss by gravitational radiation, have been supposed to be
more susceptible than other systems. But there are basically no systems
except, e.g., low-mass CVs in or below the period gap that are driven
mainly by gravitational braking. Higher mass systems are believed
to be driven mainly by magnetic breaking. For a high braking rate,
e.g., according to  \citet{Verbunt-Zwaan},
CVs with unevolved main sequence stars become unsusceptible to the
irradiation instability except for the most massive systems because
\( \tau _{\text {d}} \) becomes too small in comparison to \( \tau _{\text {ce}} \)
\cite{Ritter95, Ritter96b, Ritter00}. For giants magnetic braking
is believed to be ineffective and \( \tau _{\text {ce}} \) is small
compared to \( \tau _{\text {d}}\approx \tau _{\text {nuc}} \). Therefore,
such systems and also CVs with highly evolved MS stars, which have
a short \( \tau _{\text {ce}} \), have been supposed to be a priori
more susceptible to the irradiation instability than CVs with unevolved
MS stars \cite{King97}.

But the last term of order of several \( \frac{H_{\text {P}}}{R_{2}} \)
in (\ref{Eq:criterion:Final:1}) can be neglected only if\begin{equation}
\frac{H_{\text {P}}}{R_{2}}\ll \frac{\tau _{\text {ce}}}{\tau _{\text {d}}}.
\end{equation}
As soon as \( \tau _{\text {ce}} \) gets small (as for main sequence
stars of \( {\sim 1M_{\sun }} \)) or \( \frac{H_{\text {P}}}{R_{2}} \)
is no longer very small (as for giants), this term must be taken into
account. Obviously, the range of \( \left\langle f_{\text {irr}}\right\rangle  \)
in which\begin{equation}
s'>\frac{H_{\text {P}}}{R_{2}}\delta 
\end{equation}
 can be fulfilled, is an interval \( \left[ \left\langle f_{\text {irr,min}}\right\rangle ,\left\langle f_{\text {irr,max}}\right\rangle \right]  \)
because \( s' \) vanishes for \( \left\langle f_{\text {irr}}\right\rangle \rightarrow 0 \)
and \( \left\langle f_{\text {irr}}\right\rangle \rightarrow \infty  \).
Thus, mass transfer cycles can only occur in a limited range of \( \left\langle f_{\text {irr}}\right\rangle  \).
This range, that is predicted by the analytical model, has to be checked
by numerical computations with full stellar models.

\kommentar{It will be discussed in the following sections, to what
extent this range, which is predicted by the analytical model, is
also valid, if full stellar models are taken into account instead
of bipolytropes.}

In this section we have derived a more concise and, in some sense,
more general, expression (\ref{Eq:criterion:Final:2}), respectively
(\ref{Eq:criterion:Final:1}) for the onset of instability than the
corresponding Eq.~{[}45{]}, respectively {[}60{]} of 
\citet{King96}. We have also provided an explicit formula (\ref{Eq:ratio:timescales:numeric})
for \( \frac{\tau _{\text {ce}}}{\tau '_{\text {d}}} \) which we
will use in Sect.~\ref{Chap:Numerics} to discuss our numerical results.
An important result is that the additional term of several \( \frac{H_{\text {P}}}{R_{2}} \)
in (\ref{Eq:criterion:Final:1}) provides an upper limit for \( \tau _{\text {d}} \),
i.e., a lower limit for \( -\overline{\dot{M}}_{2} \). Hence, for
a sufficiently low secular mass transfer rate every system becomes
stable. We will show in Sect.~\ref{Sec:Numerics:CVs:Giants} that
this limit can become important for CVs with giant donors.

\section{The binary evolutionary code}

\label{Chap:Code}

We use the stellar evolutionary code of  \citet{Schlattl97, Schlattl99}
which goes back to the code of  \citet{Hofmeister}.
The code uses a special grid point algorithm \cite{Wagenhuber94},
nuclear reaction rates of  \citet{Caughlan85, Adelberger98},
the mixing length theory of  \citet{Vitense58, Cox-Giuli},
the OPAL opacities \cite{Iglesias96}, and for the outer layers opacities
by  \citet{Alexander94} and additional data
from P. H. Hauschildt and J. W. Ferguson (priv. comm.). In order to
avoid numerical instabilities in calculating mass transfer it was
necessary to implement an equation of state which is smooth in the
independent variables \( P \), \( T \), and chemical composition
over the whole range of application. In practice, we use the equation
of state of  \citet{Saumon95} with data provided
by I. Baraffe (priv. comm.), and particularly the equation of state
of  \citet{Pols95}. Furthermore, we developed
an implicit algorithm for the treatment of mass transfer similar to
the method described by  \citet{Benvenuto03}.
For more details about our code see  \citet{Buening-PHD}.
For determining \( f_{\text {int}}\left( f_{\text {irr}}\right)  \)
we use results given by  \citet{Hameury97}
and additional data from J.-M. Hameury (priv. comm.).

\section{Numerical results}

\label{Chap:Numerics}

To destabilize the stationary mass transfer the ratio of \( \tau _{\text {ce}} \)
and \( \tau '_{\text {d}} \) has to be sufficiently small, at least
less than the maximum of \( s' \) which is \( {\sim 0.1} \). As
can be seen from (\ref{Eq:ratio:timescales:numeric}), especially
for realistic values of \( \alpha \eta <1 \) the ratio \( \frac{R_{1}}{R_{2}} \)
has to be sufficiently small for typical system parameters which is
basically the case for compact binaries only. Since the system has
to be dynamically and thermally stable and since the donor star has
to maintain a deep convective envelope, essentially only CVs and LMXBs
are eligible.

\subsection{Cataclysmic variables}

\subsubsection{Unevolved MS donors}

\begin{figure}[htb]
\centering
\resizebox{\hsize}{!}{\includegraphics{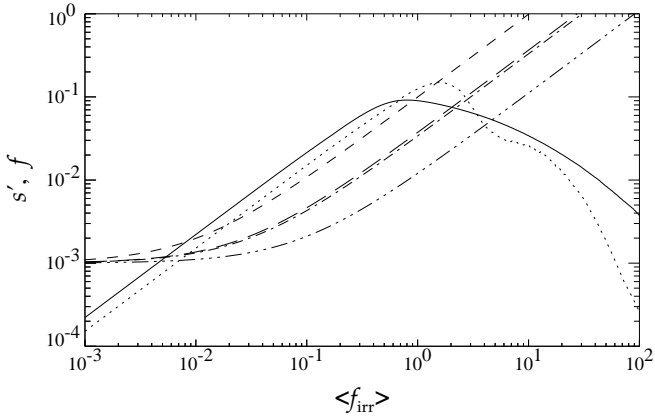}}
\caption{\( s' \) for a CV with an unevolved \( 0.5\, M_{\sun } \)
MS star (no.~2 of table~\ref{Table:SingleStars}) and an \( 0.8\, M_{\sun } \)
white dwarf for the point source and the constant flux model (solid
and dotted line). \( s' \) is compared to \( f \) for models \( 1-4 \)
listed in table \ref{Table:Binaries} (dot-dash-dotted, long-dashed,
dash-dotted and short-dashed line, respectively). The mass transfer
is unstable and undergoes cycles if \( s'>f \) which is only the
case in a certain interval \( \left[ \left\langle f_{\text {irr,min}}\right\rangle ,\left\langle f_{\text {irr,max}}\right\rangle \right]  \)
whose limits depend on \( \alpha  \). Otherwise, the system is stable.}
\label{Fig:CV:ZAMS:sf}
\end{figure}

Exemplarily for a wide range of CVs with unevolved donor stars Fig.~\ref{Fig:CV:ZAMS:sf}
shows the left-hand and also the right-hand side of (\ref{Eq:criterion:Final:1}),
namely \( s' \) and\begin{equation}
f:=\delta \frac{H_{\text {P}}}{R_{2}}+\frac{\tau _{\text {ce}}}{\tau '_{\text {d}}}
\end{equation}
for a CV with an \( 0.8\, M_{\sun } \) white dwarf and an unevolved
\( 0.5\, M_{\sun } \) MS secondary using system parameters listed
in table \ref{Table:Binaries}. Low-mass MS stars have \( \frac{H_{\text {P}}}{R_{2}}\sim 10^{-4} \)
and \( \delta \sim n+5\approx 10 \). This leads to an almost constant
contribution \( \sim 10^{-3} \) to \( f \) while \( \frac{\tau _{\text {ce}}}{\tau '_{\text {d}}} \)
is almost linear in \( \left\langle f_{\text {irr}}\right\rangle  \).
The most significant uncertainty in the values of \( f \) is due
to the undetermined value of \( \alpha  \) (short-dashed vs. dash-dotted
line in Fig.~\ref{Fig:CV:ZAMS:sf}) while the influence of \( q \)
(long-dashed vs. dash-dotted line) is negligible. Only \( \bar{\eta } \)
(dot-dash-dotted vs. dash-dotted line) has a noticable influence on
\( f \) in this case since \( \zeta _{\text {s}}-\zeta _{\text {R}} \)
is much smaller in the conservative case. Systems with non-conservative
mass transfer (\( \bar{\eta }<1 \) and \( \nu =q \)) are systematically
more stable than systems with conservative mass transfer \( \left( \bar{\eta }=1\right)  \).
In the following discussion we will restrict ourselves to \( \eta =1 \)
and \( \bar{\eta }=0 \) since this case is more realistic than \( \bar{\eta }=1 \)
\cite{Hameury89, Prialnik95}. In general, the amplitude of the mass
transfer cycles increases with the difference between \( s' \) and
\( f \) for any given \( \left\langle f_{\text {irr}}\right\rangle  \).

\begin{table}
\centering

\begin{tabular}{|c|c|c|c|c|c|c|}
\hline 
Model No.&
\( M_{1}/M_{\sun } \)&
q&
\( \bar{\eta } \)&
\( \nu  \)&
\( \zeta _{R} \)&
\( \alpha  \)\\
\hline
\hline 
1&
0.8&
\( \vphantom {\sqrt{\frac{5}{8}}}\frac{5}{8} \)&
1&
-&
-0.35&
0.3\\
\hline 
2&
1.2&
\( \vphantom {\sqrt{\frac{5}{8}}}\frac{5}{12} \)&
0&
-&
-1.20&
0.3\\
\hline 
3&
0.8&
\( \vphantom {\sqrt{\frac{5}{8}}}\frac{5}{8} \)&
0&
\( q \)&
-0.89&
0.3\\
\hline 
4&
0.8&
\( \vphantom {\sqrt{\frac{5}{8}}}\frac{5}{8} \)&
0&
\( q \)&
-0.89&
0.1\\
\hline
\end{tabular}

\caption{\label{Table:Binaries} System parameters of CVs containing an \protect\( 0.5\, M_{\sun }\protect \)
donor as shown in Fig.~\ref{Fig:CV:ZAMS:sf}.}
\end{table}

As can be seen from Fig.~\ref{Fig:CV:ZAMS:sf}, CVs with unevolved
donors are the more stable, the smaller the irradiation efficiency
\( \alpha  \) is. Such systems can undergo mass transfer cycles if
\( \alpha \gtrsim 0.1 \). For, e.g., \( \alpha =0.3 \) the {}``island
of instability{}'', which is defined by \( s'>f \), ranges from
\( \left\langle f_{\text {irr}}\right\rangle \approx 5\cdot 10^{-3} \)
to \( \left\langle f_{\text {irr}}\right\rangle \approx 2 \). Using
(\ref{Eq:ratio:timescales:numeric}) is it possible to compute the
corresponding ratio for \( \tau _{\text {ce}} \) and \( \tau '_{\text {d}} \):\begin{equation}
1.5\cdot 10^{-4}\lesssim \frac{\tau _{\text {ce}}}{\tau '_{\text {d}}}\lesssim 7\cdot 10^{-2}.
\end{equation}
The timescale ratio must be within this range to destabilize the stationary
mass transfer. With \begin{equation}
\label{Eq:Relation:tau_driving}
\tau '_{\text {d}}=\frac{\zeta _{\text {e}}-\zeta _{\text {R}}}{\zeta _{\text {s}}-\zeta _{\text {R}}}\tau _{\text {d}},
\end{equation}
which can be obtained from (\ref{Eq:Def:tau-d-prime}) and (\ref{Eq:K:stationary}),
and \( \tau _{\text {ce}}\sim 4.5\cdot 10^{7}\, \text {yr} \) for
an unevolved \( 0.5\, M_{\sun } \) MS star, we obtain a condition
for the driving timescale \( \tau _{\text {d}} \):\begin{equation}
\label{Eq:Limit:CV:unevolved}
3\cdot 10^{8}\, \text {yr}\lesssim \tau _{\text {d}}\lesssim 10^{11}\text {yr}.
\end{equation}
If the evolution of the system is driven by gravitational braking
only, i.e., \( \tau _{\text {d}}=\frac{\tau _{\text {J}}}{2}\approx 4\cdot 10^{9}\, \text {yr} \),
then the system must undergo mass transfer cycles. Gravitational braking
provides a physical upper limit for the driving timescale. Since the
lower limit in (\ref{Eq:Limit:CV:unevolved}) is given by about \( \frac{1}{10} \)
of the gravitational braking timescale, we can expect the occurrence
of the irradiation instability between \( \dot{J}=\dot{J}_{\text {grav}} \)
and \( \dot{J}\approx 10\dot{J}_{\text {grav}} \) which is roughly
confirmed by our numerical models.

\begin{figure}[htb]
\centering
\resizebox{\hsize}{!}{\includegraphics{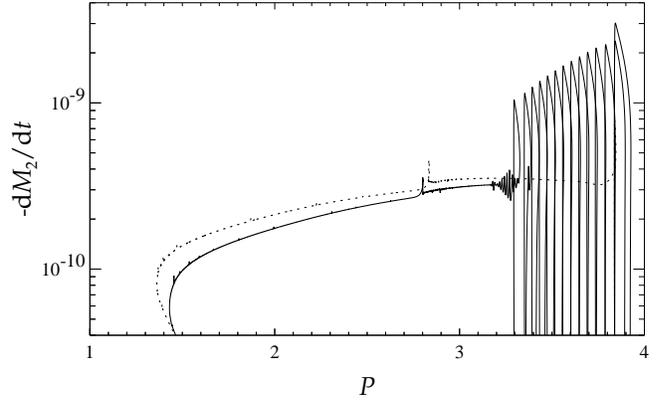}}
\caption{Mass transfer rate \( -\dot{M}_{2}\, \left[ M_{\sun }/\text {yr}\right]  \)
from a slightly evolved \( 0.5\, M_{\sun } \) MS star with an age
of \( {\sim 10^{10}\, \text {yr}} \) and \( X_{\text {c}}\approx 0.62 \)
onto an \( 0.8\, M_{\sun } \) white dwarf for the point source model
as function of orbital period \( P\, \left[ \text {hr}\right]  \).
System parameters: \( \bar{\eta }=0 \), \( \nu =q \), \( \alpha =0.3 \),
and \( \dot{J}=5\dot{J}_{\text {grav}} \) (solid line). The dotted
line shows the mass transfer rate for the same system without irradiation
feedback. Both computations have been stopped beyond the minimum period
at \( M_{2}=0.04\, M_{\sun } \).}
\label{Fig:CV:ZAMS:Evol}
\end{figure}

\begin{figure}[htb]
\centering
\resizebox{\hsize}{!}{\includegraphics{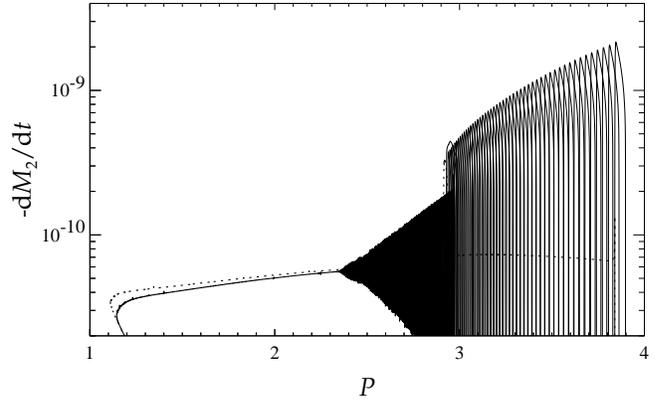}}
\caption{Same as Fig.~\ref{Fig:CV:ZAMS:Evol} but for gravitational
braking only.}
\label{Fig:CV:ZAMS:Evol:grav}
\end{figure}

Fig.~\ref{Fig:CV:ZAMS:Evol} shows a CV evolution for initial parameters
of model no.~3 of table~\ref{Table:Binaries} using the point source
model and an ad hoc braking rate of \( \dot{J}=5\dot{J}_{\text {grav}} \).
The donor, a slightly evolved MS star, fills its Roche lobe at an
orbital period of about \( 4\, \text {hr} \) and starts to transfer
mass undergoing cycles. At about \( 3.2\, \text {hr} \) the system
leaves the {}``island of instability{}'' and becomes stable. The
small spike in the mass transfer rate at about \( 2.8\, \text {hr} \)
is caused by the star becoming fully convective. For gravitational
braking only as shown in Fig.~\ref{Fig:CV:ZAMS:Evol:grav} the phase
of mass transfer cycles does not end until the system is in the middle
of the period gap. The transition, when the star becomes fully convective,
coincides with a drop in the peak mass transfer rate in this case.

The secular mass transfer rate is lower when taking irradiation into
account (dotted versus solid line in Figs.~\ref{Fig:CV:ZAMS:Evol}
and~\ref{Fig:CV:ZAMS:Evol:grav}) since the irradiated star has a
systematically larger radius. As a consequence, the minimum period
increases, in the case of pure gravitational braking by up to \( 2 \)
minutes, depending on \( \alpha  \). This is of the same order as
the correction which is caused by geometrical distortion of the mass
losing star by the Roche potential \cite{Baraffe02}, and this is
still not sufficient to explain the observed orbital minimum period
of \( \sim 78\, \text {min} \) \cite{Ritter-Katalog03} if only gravitational
braking is taken into account.

\begin{figure}[htb]
\centering
\resizebox{\hsize}{!}{\includegraphics{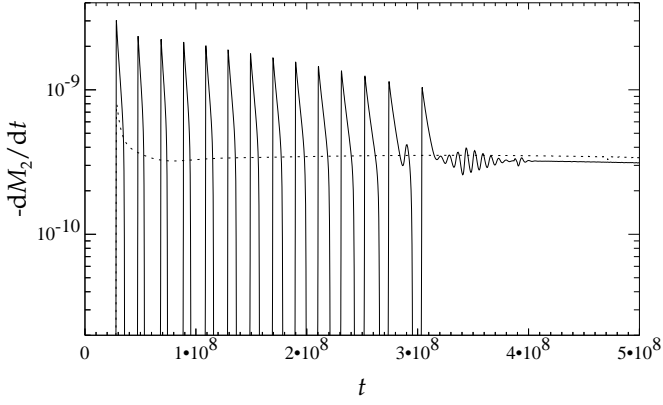}}
\caption{Mass transfer rate \( -\dot{M}_{2}\, \left[ M_{\sun }/\text {yr}\right]  \)
as function of time \( t\, \left[ \text {yr}\right]  \) for the same
system as in Fig.~\ref{Fig:CV:ZAMS:Evol}. The computation stops
at \( M_{2}\approx 0.34\, M_{\sun } \).}
\label{Fig:CV:ZAMS:Evol:2}
\end{figure}

Fig.~\ref{Fig:CV:ZAMS:Evol:2} shows the phase of mass transfer cycles
as function of time for the same system as in Fig.~\ref{Fig:CV:ZAMS:Evol}.
Since the eigenvalues at the fixed point are of the form \( \lambda \pm \text {i}\omega  \),
the frequency \( \omega  \) of the small amplitude oscillations at
the end of the unstable phase can be computed from linear stability
analysis using (\ref{Eq:DF}) and (\ref{Eq:dK-dy:2}):\begin{equation}
\label{Eq:MT:omega}
\omega =\sqrt{\det \vec{D}\vec{F}}\approx \sqrt{\frac{R_{2}}{H_{\text {P}}}\frac{1}{\tau '_{\text {d}}}\frac{\delta }{\tau _{\text {ce}}}}.
\end{equation}
For the system considered we have \( \frac{H_{\text {P}}}{R_{2}}\approx 10^{-4} \),
\( \delta \approx 10 \) and \( \tau '_{\text {d}}\approx \tau _{\text {d}} \).
The period of the low amplitude mass transfer cycles is therefore
given by\begin{equation}
\tau =\frac{2\pi }{\omega }\approx 0.02\, \sqrt{\tau _{\text {ce}}\tau _{\text {d}}}.
\end{equation}

\begin{figure}[htb]
\centering
\resizebox{\hsize}{!}{\includegraphics{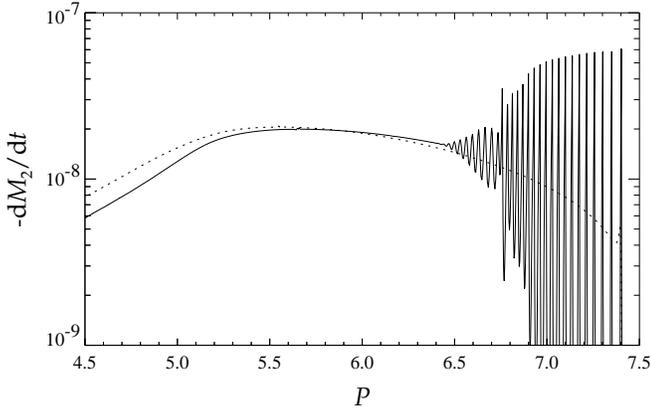}}
\caption{Mass transfer rate \( -\dot{M}_{2}\, \left[ M_{\sun }/\text {yr}\right]  \)
from a slightly evolved \( 1\, M_{\sun } \) MS star (\( X_{\text {c}}\approx 0.62 \))
onto a \( 0.8\, M_{\sun } \) white dwarf for the point source model
as function of orbital period \( P\, \left[ \text {hr}\right]  \).
System parameters: \( \bar{\eta }=0 \), \( \nu =q \), \( \alpha =0.3 \),
and strong braking according to  \citet{Verbunt-Zwaan}.
The dotted line shows the evolution without irradiation feedback.}
\label{Fig:CV:ZAMS:Evol:h}
\end{figure}

For higher braking rates, e.g., according to 
\citet{Verbunt-Zwaan}, mass transfer is stable. Only CVs with more
massive donor stars can undergo mass transfer cycles for such a high
braking rate. Fig.~\ref{Fig:CV:ZAMS:Evol:h} shows the evolution
of a CV with an initially \( 1\, M_{\sun } \) MS donor. The stationary
mass transfer rate becomes stable when the mass of the convective
envelope \( M_{\text {ce}} \) and therefore \( \tau _{\text {ce}} \)
becomes large enough  \citep[see also][]{Ritter00}.

\begin{figure}[htb]
\centering
\resizebox{\hsize}{!}{\includegraphics{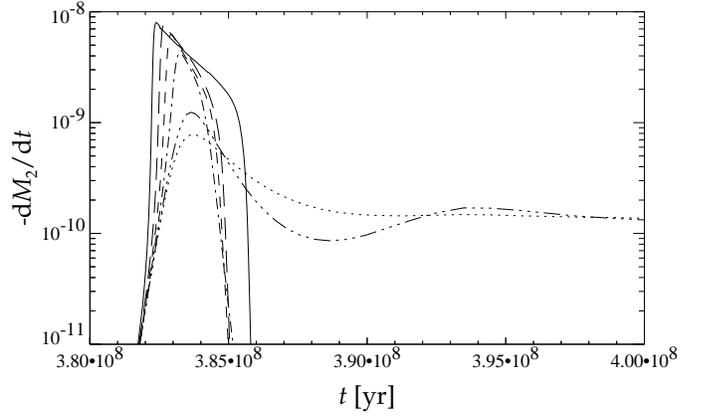}}
\caption{Mass transfer rate \( -\dot{M}_{2}\, \left[ M_{\sun }/\text {yr}\right]  \)
from an unevolved \( 0.5\, M_{\sun } \) MS star onto an \( 0.8\, M_{\sun } \)
white dwarf for conservative mass transfer \( \left( \bar{\eta }=1\right)  \),
\( \dot{J}=\dot{J}_{\text {grav}} \), and the constant flux model
for \( \alpha =0.02 \), \( 0.04 \), \( 0.1 \), \( 0.2 \), \( 0.4 \),
and \( 1.0 \) (dotted, dot-dash-dotted, dash-dotted, short-dashed,
long-dashed, and solid line, respectively) as function of time \( t\, \left[ \text {yr}\right]  \)
.}
\label{Fig:single-cycle}
\end{figure}

The dependence of the temporal evolution of mass transfer cycles on
the efficiency parameter \( \alpha  \) is shown in fig.~\ref{Fig:single-cycle}.
The duration of the high state decreases with decreasing \( \alpha  \)
until the mass transfer rate becomes sinusoidal with a frequency \( \omega  \)
as given in (\ref{Eq:MT:omega}). For even weaker feedback the oscillations
are damped or even vanish completely.

\subsubsection{Highly evolved MS donors}

The timescale of central hydrogen burning for low-mass MS stars is
much longer than a Hubble time. One way of accounting for evolved
low-mass donors in CVs is prior thermal timescale mass transfer from
an evolved, initially more massive star. In this case the mass losing
star is more massive to such an extent that the system is initially
thermally unstable (\( \zeta _{\text {e}}-\zeta _{\text {R}}<0 \))
and the mass transfer occurs on a thermal timescale \cite{KR-Cygnus, Schenker02}.

\begin{figure}[htb]
\centering
\resizebox{\hsize}{!}{\includegraphics{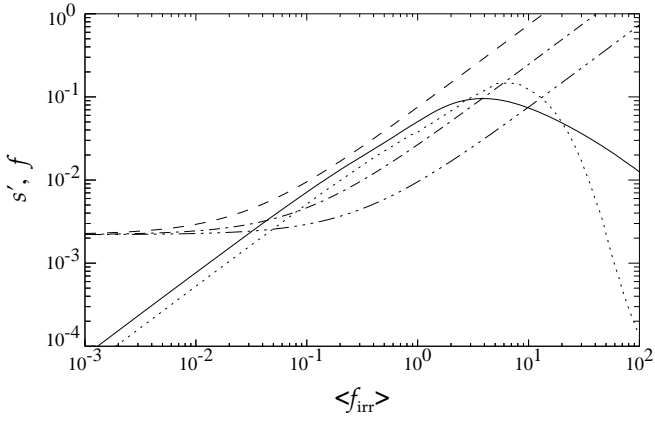}}
\caption{\( s' \) for a CV with an highly evolved \( 0.45\, M_{\sun } \)
remnant of thermal timescale mass transfer (no.~4 of table~\ref{Table:SingleStars})
and an \( 0.86\, M_{\sun } \) white dwarf for the point source and
constant flux model (solid and dotted line). \( s' \) is compared
to \( f \) for \( \bar{\eta }=0.25 \) and \( \alpha =0.1 \), \( 0.3 \),
and \( 1.0 \) (short-dashed, dash-dotted, and dot-dash-dotted line,
respectively).}
\label{Fig:CV:RTTMT:sf}
\end{figure}

Although the bipolytrope model is formally valid only for chemically
homogeneous stars we tentatively apply it also to chemically evolved
donor stars. As an example fig.~\ref{Fig:CV:RTTMT:sf} shows \( s' \)
and \( f \) for an \( 0.45\, M_{\sun } \) remnant of thermal timescale
mass transfer with a central hydrogen abundance of \( X_{\text {c}}=0.05 \).
It is the core of an initially more massive MS star of \( 1.6\, M_{\sun } \).
According to (\ref{Eq:Def:tau_ce}) \( \tau _{\text {ce}} \) is less
than \( \frac{1}{10} \) of \( \tau _{\text {ce}} \) of an unevolved
MS star of the same mass, and its radius is larger by \( \sim 70\% \).
Due to its deeper superadiabatic convection zone the maximum of \( s' \)
is shifted to higher values of \( \left\langle f_{\text {irr}}\right\rangle  \)
and due to its larger radius \( f \) is also shifted to larger values
of \( \left\langle f_{\text {irr}}\right\rangle  \), but in total
the difference to the case with an unevolved donor discussed before
is comparatively small. As can be seen from fig.~\ref{Fig:CV:RTTMT:sf},
this particular system should be stable for \( \alpha =0.1 \) but
can undergo cycles for \( \alpha =0.3 \). When going to smaller donor
masses or lower braking rates during the initial evolution also remnants
of thermal timescale mass transfer can become unstable for \( \alpha \gtrsim 0.1 \)
as CVs with unevolved donors do. It seems that CVs with an evolved
donor are not more susceptible to mass transfer cycles than CVs with
unevolved donors (if at all, it is the other way around). Rather they
are susceptible at a shorter driving timescale \( \tau _{\text {d}} \).

\begin{figure}[htb]
\centering
\resizebox{\hsize}{!}{\includegraphics{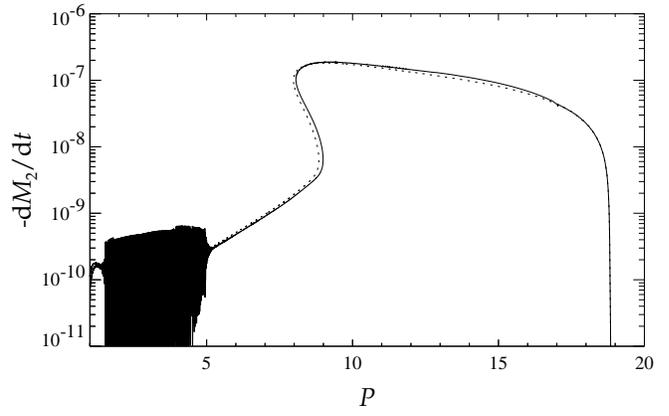}}
\caption{Mass transfer rate \( -\dot{M}_{2}\, \left[ M_{\sun }/\text {yr}\right]  \)
from an evolved initially \( 1.6\, M_{\sun } \) MS star (no.~4 of
table~\ref{Table:SingleStars}) onto an initially \( 0.6\, M_{\sun } \)
white dwarf for the point source model as function of orbital period
\( P\, \left[ \text {hr}\right]  \). System parameters: \( \bar{\eta }=0.25 \),
\( \nu =q \), \( \alpha =0.3 \), and strong braking according to
 \citet{Verbunt-Zwaan}. The dotted line shows
the evolution without irradiation feedback. The onset of mass transfer
cycles occurs at \( M_{1}\approx 0.91\, M_{\sun } \) and \( M_{2}\approx 0.26\, M_{\sun } \).}
\label{Fig:CV:RTTMT:Evol}
\end{figure}

As for unevolved CVs it is also possible to determine the range of
\( \tau _{\text {d}} \) where mass transfer cycles should occur,
e.g., if \( \alpha =0.3 \):\begin{equation}
3.5\cdot 10^{7}\, \text {yr}\lesssim \tau _{\text {d}}\lesssim 3.5\cdot 10^{9}\, \text {yr}.
\end{equation}
According to this result such a system should undergo mass transfer
cycles even for rather high braking rates. However, fig.~\ref{Fig:CV:RTTMT:Evol}
shows\footnote{%
To perform the evolutionary computations shown in figs.~\ref{Fig:CV:RTTMT:Evol}
and \ref{Fig:LMXB:RTTMT:Evol} from the initial turn-on of mass transfer
to the minimum period without any discontinuity we used opacity tables
for solar composition in the outer layers. Comparisons with computations
using correct Helium and CNO abundances for the final mass transfer
phase indicate that, though using {}``incorrect{}'' opacity tables
in that evolutionary stage leads to a different stellar radius, it
does not affect the mass transfer cycles significantly.
} that this is not the case, when the donor star reaches \( M_{2}=0.45\, M_{\sun } \)
at an orbital period of \( \sim 8\, \text {hr} \). Instead, the system
becomes unstable at a donor mass much lower than expected, i.e., at
about \( 0.26\, M_{\sun } \). The reason for this discrepancy is
probably that the bipolytrope model underestimates the timescale \( \tau _{\text {ce}} \)
on which the stellar radius changes if irradiation is changed. Numerical
computations indicate that \( \tau _{\text {ce}} \) from (\ref{Eq:Def:tau_ce})
is too large by a factor of \( \sim 1.4 \) for a fully convective
\( 0.3\, M_{\sun } \) MS star whereas it is too small by \( \sim 1.6 \)
for an \( 0.5\, M_{\sun } \) MS star and even more so for more massive
or more evolved stars. As a result evolved systems are significantly
more stable than previously expected.

\subsubsection{Giant donors}

\label{Sec:Numerics:CVs:Giants}

\begin{figure}[htb]
\centering
\resizebox{\hsize}{!}{\includegraphics{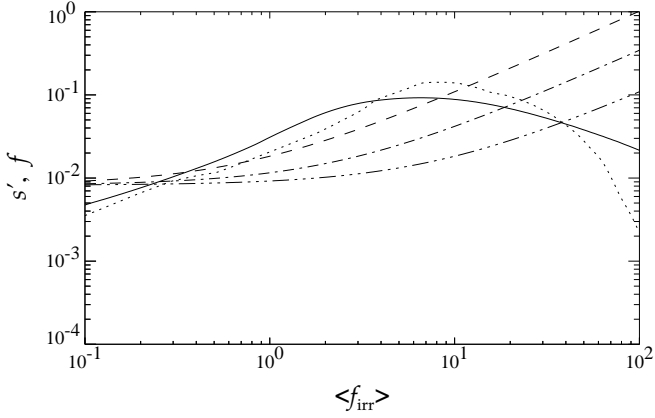}}
\caption{\( s' \) for a CV with an \( 0.8\, M_{\sun } \) giant (no.~6
of table~\ref{Table:SingleStars}) and an \( 1.2\, M_{\sun } \)
white dwarf for the point source and the constant flux model (solid
and dotted line). \( s' \) is compared to \( f \) for \( \bar{\eta }=0 \)
and \( \alpha =0.1 \), \( 0.3 \), and \( 1.0 \) (short-dashed,
dash-dotted, and dot-dash-dotted line, respectively).}
\label{Fig:CV:GIANT:sf}
\end{figure}

Another group of CVs are systems containing a giant or subgiant donor.
Fig.~\ref{Fig:CV:GIANT:sf} shows the stability diagram for a CV
with a giant donor of \( 0.8\, M_{\sun } \) (model no.~6 of table~\ref{Table:SingleStars}).
Since \( \tau _{\text {ce}} \) according to (\ref{Eq:Def:tau_ce})
gives an unreliable estimate for the thermal timescale of the convective
envelope of giants, we have used the timescale \( \tau _{\text {ce,U}}\sim 6\cdot 10^{3}\, \text {yr} \)
which is defined as the thermal energy \( U \) of the convective
envelope divided by the stellar luminosity. The timescale on which
the giant radius changes, when irradiation is changed, is even larger
than \( \tau _{\text {ce,U}} \) by \( \sim 2 \). Giants typically
have \( \frac{H_{\text {P}}}{R_{2}}\sim 10^{-2} \) and \( \delta \sim n+5\approx 2 \).
In our case for \( \alpha =0.3 \) the {}``island of instability{}''
ranges from \( 0.25 \) to \( 20 \) in \( \left\langle f_{\text {irr}}\right\rangle  \).
This means the ratio of \( \tau _{\text {ce}} \) and \( \tau '_{\text {d}} \)
must be in the range\begin{equation}
8.5\cdot 10^{-4}\lesssim \frac{\tau _{\text {ce}}}{\tau '_{\text {d}}}\lesssim 7\cdot 10^{-2},
\end{equation}
or taking into account (\ref{Eq:Relation:tau_driving})\begin{equation}
5.5\cdot 10^{-4}\lesssim \frac{\tau _{\text {ce}}}{\tau _{\text {d}}}\lesssim 4\cdot 10^{-2}.
\end{equation}
Using analytical approximations given by 
\citet[ eq.~43]{Ritter99} and (\ref{Eq:Relation:tauM_taud}) we can
write\begin{equation}
\label{Eq:ratio:giant}
\frac{\tau _{\text {ce}}}{\tau _{\text {d}}}\approx -3\cdot 10^{-4}\frac{M_{2}-M_{\text {c}}}{M_{2}}\left( \frac{M_{2}}{M_{\sun }}\right) ^{2}\left( \frac{0.25\, M_{\sun }}{M_{\text {c}}}\right) ^{6}
\end{equation}
for Pop. I giants and a similar expression for Pop. II giants. The
temporal evolution of the core mass \( M_{\text {c}} \) is given
by\begin{equation}
M_{\text {c}}=M_{\text {c,i}}\left( 1-\frac{t}{t_{\infty }}\right) ^{-\frac{1}{7}}.
\end{equation}
Here, \( M_{\text {c,i}} \) denotes the initial core mass and \( t_{\infty } \)
the nuclear timescale, i.e., the time when the core mass formally
becomes infinite. Almost all values on the right-hand side of (\ref{Eq:ratio:giant})
are less than unity. We have \( M_{\text {c}}<M_{2} \) and for thermally
stable systems typically \( M_{2}<M_{\sun } \). Accordingly, \( \frac{\tau _{\text {ce}}}{\tau _{\text {d}}} \)
is too small for destabilizing the system except if \( M_{\text {c}}<0.25\, M_{\sun } \)
and \( M_{\text {c}}\ll M_{2} \).

For the system shown in fig.~\ref{Fig:CV:GIANT:sf} Eq.~(\ref{Eq:ratio:giant})
yields:\begin{equation}
\frac{\tau _{\text {ce}}}{\tau _{\text {d}}}\approx 5\cdot 10^{-5}.
\end{equation}
 In fact, to destabilize the system \( \alpha  \) would have to be
larger by about an order of magnitude than the value of \( 0.3 \)
which we have chosen for this example. Since an efficiency \( \alpha >1 \)
is rather unrealistic, we conclude that CVs with extended giants can
not undergo mass transfer cycles.

There are three reasons, why CVs with giant donors are the more stable,
the more extended the giants are: First, 
\citet{King97} have neglected the last term in (\ref{Eq:criterion:Final:1})
that restricts the region of instability not only to high but also
to low braking rates. Thus, systems do not get automatically unstable
for sufficiently small braking rates. Second, the depth of the superadiabatic
convection zone grows during giant evolution which shifts the {}``island
of instability{}'' to higher \( \left\langle f_{\text {irr}}\right\rangle  \).
Hence, more evolved giants are less affected by irradiation. Third,
more evolved giants have a larger scale height \( \frac{H_{\text {P}}}{R_{2}} \)
so that their {}``island of instability{}'' is smaller than for
less evolved giants.

\subsection{Low-mass X-ray binaries}

\subsubsection{Unevolved MS donors}

As we have discussed in the last section, CVs might undergo mass transfer
cycles for \( 0.1\lesssim \alpha \lesssim 1 \) given a suitable driving
timscale. Another class of compact binary systems with thermally stable
mass transfer, i.e., small \( q \), are LMXBs. In the following we
consider only LMXBs with neutron star primaries. In principle, the
discussion for these system is analogous to the discussion of CVs.
While the radius of a white dwarf is about \( 10^{-2}\, R_{\sun } \),
the radius of a neutron star is only about \( 10\, \text {km} \).
Since for a given secondary star and driving mechanism \( \left\langle f_{\text {irr}}\right\rangle \sim \frac{\alpha }{R_{1}} \)
(cf. Eq. (\ref{Eq:ratio:timescales:numeric})) the smaller radius
of a neutron star has to be compensated by a correspondingly small
value of \( \alpha  \) for achieving a comparable value of \( \left\langle f_{\text {irr}}\right\rangle  \).
Because an efficiency that small is unlikely, 
\citet{King96, King97, Ritter00} have concluded that irradiation in
LMXBs is too strong to destabilize such systems.

\begin{figure}[htb]
\centering
\resizebox{\hsize}{!}{\includegraphics{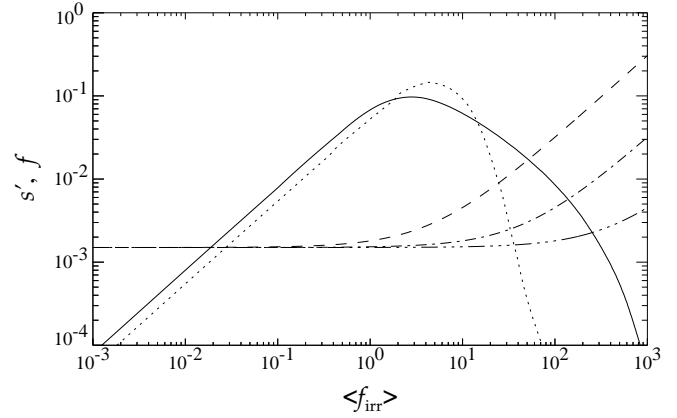}}
\caption{\( s' \) for an LMXB with an unevolved \( 0.8\, M_{\sun } \)
MS star (no.~\( 3 \) of table~\ref{Table:SingleStars}) and an
initially \( 1.4\, M_{\sun } \) NS for the point source and the constant
flux model (solid and dotted line). \( s' \) is compared to \( f \)
for \( \bar{\eta }=1 \) and \( \alpha =0.01 \), \( 0.1 \), and
\( 1.0 \) (short-dashed, dash-dotted, and dot-dash-dotted line, respectively).}
\label{Fig:LMXB:ZAMS:sf}
\end{figure}

\begin{figure}[htb]
\centering
\resizebox{\hsize}{!}{\includegraphics{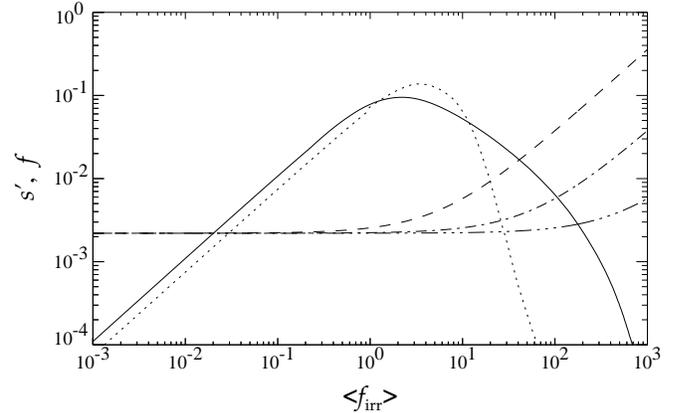}}
\caption{\( s' \) for an LMXB with an \( 0.6\, M_{\sun } \) remnant
of thermal timescale mass transfer (no.~\( 5 \) of table \ref{Table:SingleStars})
and an initially \( 1.4\, M_{\sun } \) NS for the point source and
the constant flux model (solid and dotted line). \( s' \) is compared
to \( f \) for \( \bar{\eta }=1 \) and \( \alpha =0.01 \), \( 0.1 \),
and \( 1.0 \) (short-dashed, dash-dotted, and dot-dash-dotted line,
respectively).}
\label{Fig:LMXB:RTTMT:sf}
\end{figure}

\begin{figure}[htb]
\centering
\resizebox{\hsize}{!}{\includegraphics{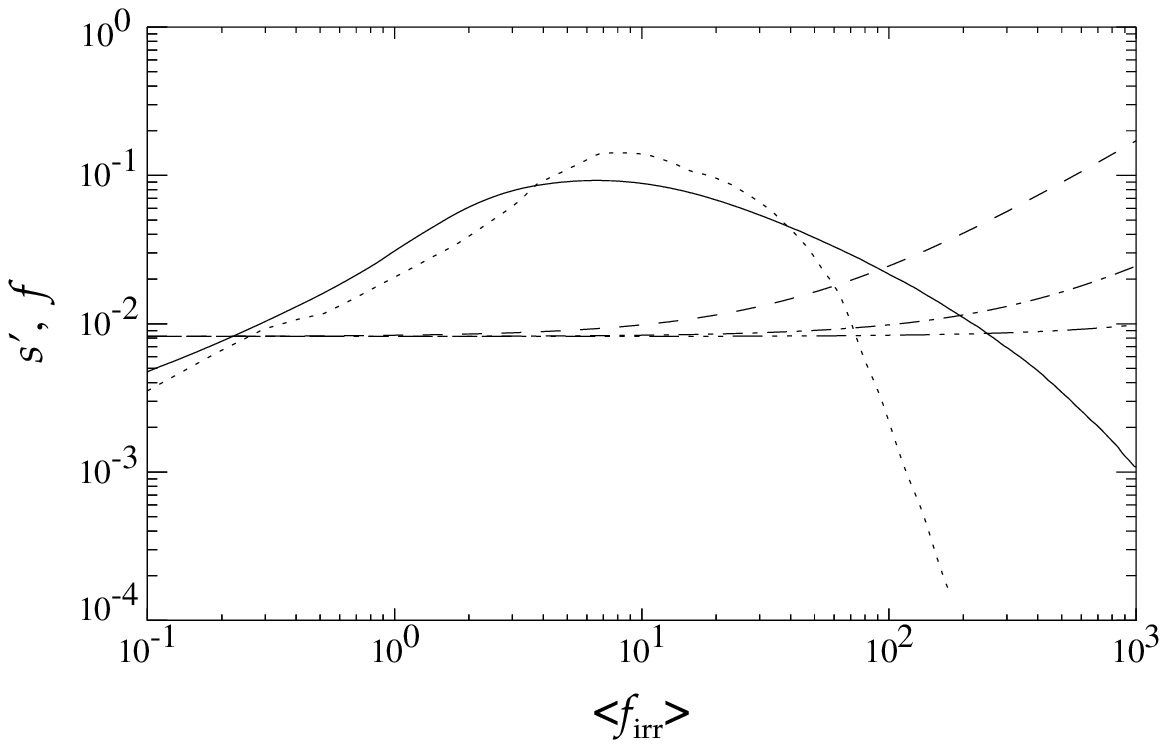}}
\caption{\( s' \) for an LMXB with an \( 0.8\, M_{\sun } \) giant
(no.~\( 6 \) of table \ref{Table:SingleStars}) and an initially
\( 1.4\, M_{\sun } \) NS for the point source and the constant flux
model (solid and dotted line). \( s' \) is compared to \( f \) for
\( \bar{\eta }=1 \) and \( \alpha =0.01 \), \( 0.1 \), and \( 1.0 \)
(short-dashed, dash-dotted, and dot-dash-dotted line, respectively).}
\label{Fig:LMXB:GIANT:sf}
\end{figure}

However, this conclusion applies only to the constant flux model but
not to the point source model. For the point source model \( s' \)
decreases much more slowly for large \( \left\langle f_{\text {irr}}\right\rangle  \)
than for the constant flux model because the surface elements near
the terminator are only partially blocked by irradiation, even for
very high fluxes due to the high angle of incidence. Therefore, the
point source model does not require \( \left\langle f_{\text {irr}}\right\rangle \sim 1 \)
to get mass transfer cycles. Instead, \( s' \) becomes larger than
\( f \) for fluxes below \( \left\langle f_{\text {irr}}\right\rangle \lesssim 10^{2}-10^{3} \)
as can be seen in fig.~\ref{Fig:LMXB:ZAMS:sf} for a LMXBs with an
unevolved donor, in fig.~\ref{Fig:LMXB:RTTMT:sf} for a LMXBs with
a remnants of thermal timescale mass transfer, and in fig.~\ref{Fig:LMXB:GIANT:sf}
for a LMXBs with a giant. Thus, LMXBs can become unstable for \( \alpha \lesssim 10^{-2}-10^{-1} \).

\begin{figure}[htb]
\centering
\resizebox{\hsize}{!}{\includegraphics{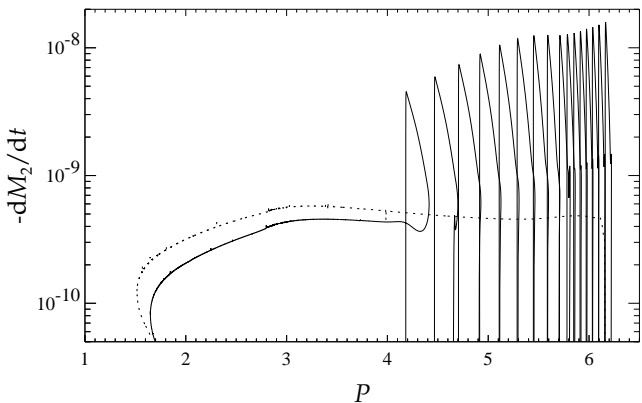}}
\caption{Mass transfer rate \( -\dot{M}_{2}\, \left[ M_{\sun }/\text {yr}\right]  \)
from an evolved (\( X_{\text {c}}\approx 0.66 \)) initially \( 0.8\, M_{\sun } \)
MS star onto an initially \( 1.4\, M_{\sun } \) neutron star for
the point source model as function of orbital period \( P\, \left[ \text {hr}\right]  \).
System parameters: \( \bar{\eta }=1 \), \( \alpha =0.01 \), and
strong braking according to  \citet{Verbunt-Zwaan}.
The dotted line shows the evolution without irradiation feedback.}
\label{Fig:LMXB:ZAMS:Evol}
\end{figure}

The {}``island of instability{}'' for the LMXB with an unevolved
donor, which is shown in fig.~\ref{Fig:LMXB:ZAMS:sf}, ranges from
\( 0.02 \) to \( 150 \) in \( \left\langle f_{\text {irr}}\right\rangle  \)
for \( \alpha =0.1 \). This corresponds to\begin{equation}
6\cdot 10^{-7}\lesssim \frac{\tau _{\text {ce}}}{\tau '_{\text {d}}}\lesssim 4.5\cdot 10^{-3}.
\end{equation}
With \( \tau _{\text {ce}}\approx 2\cdot 10^{6}\, \text {yr} \) for
an unevolved \( 0.8\, M_{\sun } \) MS star and using (\ref{Eq:Relation:tau_driving})
we get\begin{equation}
4.5\cdot 10^{8}\, \text {yr}\lesssim \tau _{\text {d}}\lesssim 3.5\cdot 10^{12}\, \text {yr}.
\end{equation}
From this we can infer that for \( \alpha =0.1 \) such a system shows
cycles for gravitational braking with \( \tau _{\text {grav}}=\frac{\tau _{\text {J}}}{2}\approx 6\cdot 10^{9}\, \text {yr} \)
and also for higher braking rates of up to \( 10\dot{J}_{\text {grav}} \).
Numerical computations show that this system is rather near to the
boundary of the unstable zone. Since the bipolytrope model underestimates
\( \tau _{\text {ce}} \) by a factor of \( \sim 2 \), the system
undergoes cycles only for braking rates up to \( 5\dot{J}_{\text {grav}} \).
Only low-amplitude sinusoidal oscillations occur for \( \alpha =0.1 \)
while for \( \alpha =10^{-2} \) the system is deep in the unstable
region as the evolution in fig.~\ref{Fig:LMXB:ZAMS:Evol} demonstrates.

\subsubsection{Highly evolved MS donors}

\begin{figure}[htb]
\centering
\resizebox{\hsize}{!}{\includegraphics{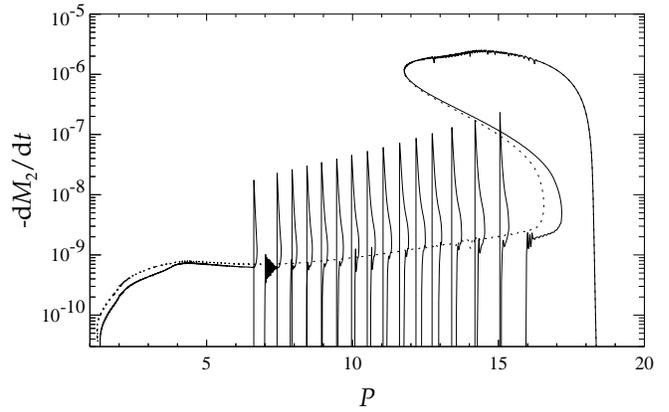}}
\caption{Mass transfer rate \( -\dot{M}_{2}\, \left[ M_{\sun }/\text {yr}\right]  \)
from an evolved, initially \( 3.0\, M_{\sun } \) MS star (no.~5
of table~\ref{Table:SingleStars}) onto an initially \( 1.4\, M_{\sun } \)
neutron star for the point source model and \( \alpha =0.1 \) as
function of orbital period \( P\, \left[ \text {hr}\right]  \). For
this computation \( \bar{\eta }=\eta =1 \) has been used as long
as the Eddington accretion rate of the neutron star \( \left( 2\cdot 10^{-8}\, \dot{M}_{\sun }/\text {yr}\right)  \)
is not exceeded, otherwise \( \bar{\eta }=\eta <1 \). The dotted
line shows the evolution of the same system without irradiation feedback.}
\label{Fig:LMXB:RTTMT:Evol}
\end{figure}

There is only a small region in the parameter space where LMXBs can
undergo significant cycles for an irradiation efficiency as large
as \( \sim 0.1 \) and for high braking rates according to  \citet{Verbunt-Zwaan}.
As an example we show in fig.~\ref{Fig:LMXB:RTTMT:sf} the evolution
of a LMXB through thermal timescale mass transfer. The {}``island
of instability{}'' for the \( 0.6\, M_{\sun } \) remnant (model
no.~5 of table~\ref{Table:SingleStars}), corresponding to an orbital
period of \( \sim 8\, \text {hr} \), is shown in fig.~\ref{Fig:LMXB:RTTMT:Evol}
and ranges from \( 0.02 \) to \( 100 \) in \( \left\langle f_{\text {irr}}\right\rangle  \),
corresponding to\begin{equation}
7\cdot 10^{-7}\lesssim \frac{\tau _{\text {ce}}}{\tau '_{\text {d}}}\lesssim 3.5\cdot 10^{-3}.
\end{equation}
With \( \tau _{\text {ce}}\approx 5\cdot 10^{5}\, \text {yr} \) the
driving timescale \( \tau _{\text {d}} \) must be between \( 2\cdot 10^{8}\, \text {yr} \)
and \( 10^{12}\, \text {yr} \) so that the system should be near
the boundary of the unstable region. This is roughly consistent with
the fact that the system becomes stable just one cycle later at \( P\sim 7\, \text {hr} \).

\subsubsection{Giant donors}

\begin{figure}[htb]
\centering
\resizebox{\hsize}{!}{\includegraphics{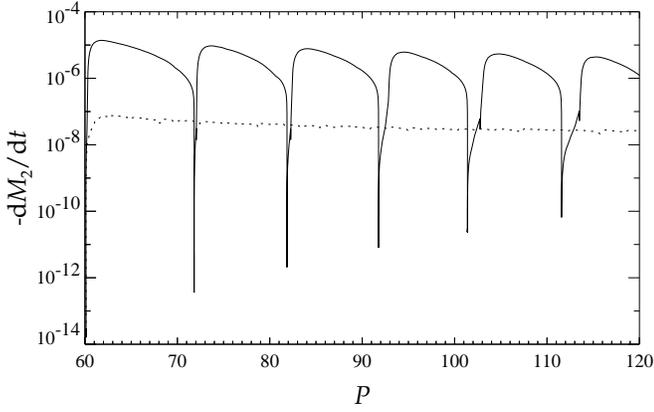}}
\caption{Mass transfer rate \( -\dot{M}_{2}\, \left[ M_{\sun }/\text {yr}\right]  \)
from an \( 0.8\, M_{\sun } \) giant (no.~6 of table~\ref{Table:SingleStars})
onto an initially \( 1.4\, M_{\sun } \) neutron star for the point
source model, \( \bar{\eta }=1 \), and \( \alpha =0.1 \) as function
of orbital period \( P\, \left[ \text {d}\right]  \). The dotted
line shows the evolution of the same system without irradiation feedback.}
\label{Fig:LMXB:GIANT:Evol}
\end{figure}

\begin{figure}[htb]
\centering
\resizebox{\hsize}{!}{\includegraphics{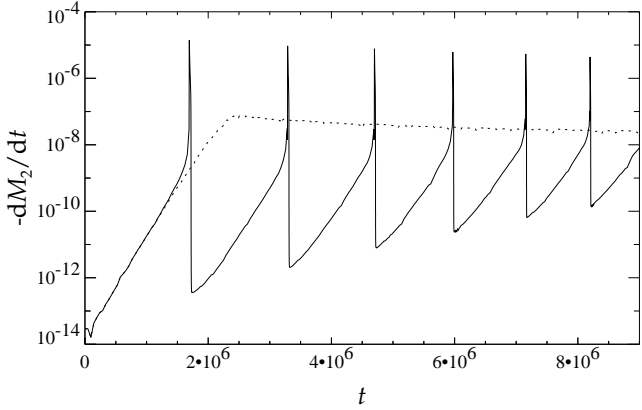}}
\caption{Same as fig.~\ref{Fig:LMXB:GIANT:Evol} but \( -\dot{M}_{2} \)
as function of time \( t\, \left[ \text {yr}\right]  \).}
\label{Fig:LMXB:GIANT:Evol:2}
\end{figure}

Unlike CVs with giant donors LMXBs with giant donors can undergo mass
transfer cycles even for very extended giants. This can be seen from
the example shown in fig.~\ref{Fig:LMXB:GIANT:sf} which illustrates
the conditions at the beginning of the evolution shown in Figs.~\ref{Fig:LMXB:GIANT:Evol}
and~\ref{Fig:LMXB:GIANT:Evol:2}. Unlike all other evolutions shown
before this system evolves to longer orbital periods because it is
driven by nuclear evolution, not loss of angular momentum.

However, taking into account, that a neutron star has a quite small
Eddington accretion rate of \( \sim 2\cdot 10^{-8}\, M_{\sun }/\text {yr} \),
this particular giant system would be stable, since the secular mass
transfer rate is above the Eddington rate, so that there would be
no irradiation feedback.

\subsection{General outline of mass transfer cycles \corr{and observational implications}}

The appearance of mass transfer cycles differs between different types
of systems. {}``Ideal{}'' outbursts as outlined by 
\citet{King97} require a very short ratio of \( \tau _{\text {ce}} \)
and \( \tau '_{\text {d}} \) which is typical for giants. They are
characterised by a very short high state at a rather high mass transfer
rate followed by a long lasting low state as shown by fig.~\ref{Fig:LMXB:GIANT:Evol:2}.

In the limit of weak irradiation feedback, as it is typically the
case for CVs with \( 0.1\lesssim \alpha \lesssim 1 \) and weak braking,
a system leaves the high state long before saturation is reached.
Only a few \( 10^{-3} \) of the stellar mass is transferred during
a cycle and the recurrence timescale is rather short. In the limit
of strong irradiation feedback, as it is \corr{often} the case for
LMXBs with \( 10^{-2}\lesssim \alpha \lesssim 0.1 \), the system
stays in the high state until saturation has been almost exactly reached.
As a consequence of this, the mass transfer rate decreases to almost
the stationary value before the system enters the low state. The duration
of the high and low state are of the same order, and in the high state
the system transfers mass just slightly above the stationary mass
transfer rate for most of the time.

\corr{Since the timescale of mass transfer cycles, i.e., the thermal
timescale of the convective envelope of the donor is by far longer
than typical observational timescales (at most of order of centuries),
it is effectively not possible to observe the evolution of a single
system from the high into the low state, or vice versa. Nevertheless,
the occurrence of these cycles would affect the observable properties
of a binary population. First, systems in the low state are more difficult
to observe than in the bright high state or may even appear as detached
systems if mass transfer ceases completely in the low state. If the
duration of the low state is longer than that of the high state, then
most of the population can be basically unobservable. Second, even
in the high state the mass transfer does not proceed at a constant
rate but declines continuously from the peak rate until the system
{}``switches{}'' into the low state. Hence, even a completely homogeneous
sample can show significant variations in the mass transfer rate.
Third, the mean mass transfer rate of the observable systems in the
high state can be significantly larger than the secular mass transfer
rate which can lead to apparent discrepancies between observations
and theoretical predictions if the occurrence of mass transfer cycles
is not taken into account.}

\corr{Therefore, mass transfer cycles could provide an explanation
for, e.g., the scatter of mass transfer rates of novalike CVs above
the period gap, the disappearance of bright systems at the upper edge
of the period gap, or discrepancies between predicted and observed
luminosities and birthrates of LMXBs as recently suggested by 
\citet{Pfahl03}.}

\subsection{Non-local effects}

\label{Sec:Numerics:nonlocal}

\corr{We have assumed that irradiation is a local effect and can be
treated for every surface element of the donor separately. If the
contribution of non-local effects (e.g., circulations) to the lateral
heat transport becomes non-negligible for sufficiently strong irradiation,
or if there is another mechanism which heats the unilluminated side
(e.g., scattering of X-rays by an extended X-ray corona), then also
the effective temperature and the intrinsic flux on the unilluminated
surface depends on \( \left\langle f_{\text {irr}}\right\rangle  \)
and differs from \( T_{0} \) and \( F_{0} \), respectively.}

\corr{Observations of the eclipsing binary AA Dor show that even in
the case of very strong irradiation an illuminated star can preserve
a cool hemisphere with \( \sim 2\, 000\, \text {K} \) while the illuminated
side is heated up to \( \sim 20\, 000\, \text {K} \) \cite{Hilditch03}.
Therefore, it seems plausible that also in the case of LMXBs at most
a small fraction of the irradiating flux is transported to the unilluminated
side. Nevertheless, for sufficiently strong irradiation, e.g., \( \left\langle f_{\text {irr}}\right\rangle \sim 10^{4} \),
even a small fraction can have a significant effect onto the unilluminated
side. We note that with the general definition of \( s \) in (\ref{Eq:Def:s})
our analytical ansatz remains valid, even if the unilluminated side
is heated up, and we formally get the same stability criterion. In
this case the shape of \( s' \) depends on the exact prescription
of those non-local effects. Here we can not make any quantitative
predictions. Nevertheless, it is plausible that \( s' \) becomes
greater at higher fluxes \( \left\langle f_{\text {irr}}\right\rangle  \)
if non-local effects are taken into account. This would mean that
our local treatment of irradiation provides a lower limit on the occurrence
of mass transfer cycles in LMXBs. An upper limit is given by the fact
that the area below the \( s' \) curve has to be less than the maximum
fraction of the surface which is affected by irradiation, i.e., less
than unity.}

\section{Summary and Conclusion}

\label{Chap:Summary}

We have described the physics of irradiation driven mass transfer
cycles by a two-dimensional system of ordinary differential equations
similar to  \citet{King96}. An additional
term of order of several times \( \frac{H_{\text {P}}}{R_{2}} \)
which reflects the usual thermal relaxation of the mass losing star,
can not be neglected, because it provides an upper limit for the driving
timescale \( \tau _{\text {d}} \), respectively a lower limit for
the timescale ratio \( \frac{\tau _{\text {ce}}}{\tau '_{\text {d}}} \),
where mass transfer cycles can occur. Taking into account this term
and also the rather large pressure scale height and the deep superadiabatic
convection zone of giants we have concluded that CVs with giants above
a certain core mass can not undergo mass transfer cycles unless the
efficiency parameter \( \alpha  \) is unexpectedly large. On the
other hand LMXBs with giants might be susceptible to the irradiation
instability as it has been predicted by  \citet{King97}.

We agree with results of  \citet{King96, Ritter00}
regarding CVs with unevolved MS stars: For high braking rates 
\citep[e.g.,][]{Verbunt-Zwaan}, as they are also required by the period
gap model  \citep[e.g.,][]{Spruit83, Kolb93},
such systems are stable except for the most massive \( \left( \sim 1\, M_{\sun }\right)  \)
donor stars. On the other hand, for low braking rates, as they are
proposed by  \citet{Sills00, Andronov03},
CVs above the period gap might undergo cycles. In this case only novalikes
are affected since dwarf novae have rather large amplitude outbursts
with a rather small duty cycle and therefore a small \( \alpha _{\text {d}} \).
Even for gravitational braking the susceptability to irradiation ceases
within the period gap so that CVs with unevolved MS stars below the
period gap are stable.

Numerical evolutionary computations have shown that the uncritical
application of the predictions of the analytical model to highly evolved
remnants of thermal timescale mass transfer does not yield good quantitative
estimates for the boundaries of the unstable region. The timescale
on which the stellar radius changes, if irradiation is changed, seems
to be significantly underestimated by the bipolytrope model so that
such systems are more stable at high braking rates than previously
expected. The most evolved systems are susceptible to irradiation
for \( 0.1\lesssim \alpha \lesssim 1 \) and orbital periods roughly
between \( 2 \) and \( 5 \) hours, or donor masses between \( 0.1 \)
and \( 0.3\, M_{\sun } \), respectively. For giants the ratio of
thermal energy of the envelope divided by the stellar luminosity seems
to give a good estimate for \( \tau _{\text {ce}} \).

Since in the point source model surface elements near the terminator
are irradiated at a high angle of incidence, they are only partially
blocked by irradiation even for high fluxes. Thus, other than expected
by  \citet{King96, King97, Ritter00} also
LMXBs with unevolved or evolved MS stars might be susceptible to the
irradiation instability. Regarding the driving timescale for such
systems basically the same restrictions apply as for CVs. LMXBs can
undergo mass transfer cycles only for \( \alpha \lesssim 0.1 \).
\corr{The contribution of non-local effects of irradiation, e.g.,
circulations, which were not taken into account by our model, might
be non-negligible for LMXBs. It seems plausible that this effect tendentially
destabilizes the mass transfer.} However, it is not clear \corr{in
all cases} whether mass transfer in LMXBs is driven by Roche lobe
overflow or by irradiation-induced winds \cite{Basko73, Basko77, Iben97}.
In the latter case our model could obviously not be applied to these
systems.

We conclude that the analytical model gives a qualitatively and also
quantitatively suitable description for the onset of mass transfer
cycles except for highly evolved remnants of thermal timescale mass
transfer. We think that the boundaries of the unstable regions can
be determined with an accuracy of a factor of \( \lesssim 2 \).

To answer the final question whether the observable CV or LMXB population
or parts of it is undergoing mass transfer cycles we need to know
two basic parameters: the driving timescale, respectively the magnetic
braking law, and the efficiency \( \alpha  \). Nevertheless, we have
shown that irradiation-driven mass transfer cycles in compact binaries
are possible for not unreasonable values of \( \alpha  \) and \( \tau _{\text {d}} \).

\begin{acknowledgements}
We thank A. Weiss and H. Schlattl for providing us with their stellar
evolutionary code, H. Schlattl for valuable support and helpful discussions
about numerics, I. Baraffe for providing high resolution EOS tables,
P. P. Eggleton for providing us with his EOS code, P. H. Hauschildt
and J. W. Ferguson for providing opacity tables, and J.-M. Hameury
for providing high resolution irradiation tables. \corr{We also thank
our referee P. Podsiadlowski for his helpful comments and suggestions.}
\end{acknowledgements}
\bibliographystyle{aa}
\bibliography{astro,klassik}

\end{document}